\documentclass[prd,amsmath,superscriptaddress,twocolumn]{revtex4}

\usepackage{hyperref}
\usepackage{ifthen}
\usepackage{amsmath}
\usepackage{amssymb}
\usepackage{graphics}
\usepackage{color,bm}

\newcommand{\Sm}{S_{\rm m}}

\newcommand{\tg}{\tilde{g}}
\newcommand{\Mpch}{\mbox{Mpc}/h}

\newcommand{\Msunh}{M_\odot /h}

\definecolor{darkgreen}{cmyk}{0.85,0.2,1.00,0.2}

\newcommand{\bq}{\begin{equation}}
\newcommand{\eq}{\end{equation}}
\newcommand{\bqa}{\begin{eqnarray}}
\newcommand{\eqa}{\end{eqnarray}}

\definecolor{darkgreen}{cmyk}{0.85,0.2,1.00,0.2}
\definecolor{purple}{cmyk}{0.5,1.0,0,0}

\newcommand{\hMpc}{h^{-1}~\text{Mpc}}

\newcommand{\absfR}{|f_{R0}|}

\newcommand{\rmd}{\ensuremath{\mathrm{d}}}

\newcommand{\rs}{r_{\rm s}}
\newcommand{\rvir}{r_{\rm vir}}
\newcommand{\rta}{r_{\rm ta}}
\newcommand{\cvir}{c_{\rm vir}}
\newcommand{\Mvir}{M_{\rm vir}}
\newcommand{\rhom}{\rho_{\rm m}}
\newcommand{\brhom}{\bar{\rho}_{\rm m}}
\newcommand{\drhom}{\delta\rhom}
\newcommand{\rhoc}{\bar{\rho}_{c}}
\newcommand{\rhos}{\rho_{\rm s}}
\newcommand{\rhoEdS}{\bar{\rho}_{\rm EdS}}
\newcommand{\sigmas}{\sigma_{\rm s}}

\newcommand{\dfR}{\delta f_R}
\newcommand{\dfRlin}{\delta f_R^{\rm lin}}
\newcommand{\dfRcham}{\delta f_R^{\rm cham}}
\newcommand{\fRlin}{f_R^{\rm lin}}
\newcommand{\fRcham}{f_R^{\rm cham}}

\newcommand{\PsiGR}{\Psi_{\rm GR}}

\begin{document}

\title{Chameleon $f(R)$ gravity in the virialized cluster}

\author{Lucas~Lombriser}
\affiliation{Institute of Cosmology \& Gravitation, University of Portsmouth, Portsmouth, PO1 3FX, UK}
\author{Kazuya~Koyama}
\affiliation{Institute of Cosmology \& Gravitation, University of Portsmouth, Portsmouth, PO1 3FX, UK}
\author{Gong-Bo~Zhao}
\affiliation{Institute of Cosmology \& Gravitation, University of Portsmouth, Portsmouth, PO1 3FX, UK}
\affiliation{National Astronomy Observatories, Chinese Academy of Science, Beijing, 100012, P.~R.~China}
\author{Baojiu~Li}
\affiliation{Institute for Computational Cosmology, Physics Department, University of Durham, South Road, Durham, DH1 3LE, UK}

\date{\today}

\begin{abstract}
Current constraints on $f(R)$ gravity from the large-scale structure are at the verge of penetrating into a region where the modified forces become nonlinearly suppressed.
For a consistent treatment of observables at these scales, we study cluster quantities produced in \emph{chameleon} and \emph{linearized} Hu-Sawicki $f(R)$ gravity dark matter $N$-body simulations. We find that the standard Navarro-Frenk-White halo density profile and the radial power law for the pseudo phase-space density provide equally good fits for $f(R)$ clusters as they do in the Newtonian scenario. We give qualitative arguments for why this should be the case. For practical applications, we derive analytic relations, e.g., for the $f(R)$ scalar field, the gravitational potential, and the velocity dispersion as seen within the virialized clusters. These functions are based on three degrees of freedom fitted to simulations, i.e., the characteristic density, scale, and velocity dispersion. We further analyze predictions for these fitting parameters from the gravitational collapse and the Jeans equation, which are found to agree well with the simulations. Our analytic results can be used to consistently constrain chameleon $f(R)$ gravity with future observations on virialized cluster scales without the necessity of running a large number of simulations.
\end{abstract}

\maketitle

\section{Introduction}

Modifications of gravity can serve as an alternative explanation to the dark energy paradigm for the late-time accelerated expansion of our Universe.
Here, we specialize to $f(R)$ gravity, where the Einstein-Hilbert action is supplemented with a free nonlinear function $f(R)$ of the Ricci scalar $R$~\cite{buchdahl:70}.
It has been shown that such models can reproduce the cosmic acceleration without invoking dark energy~\cite{carroll:03, nojiri:03, capozziello:03}.
However, they also produce a stronger gravitational coupling and enhance the growth of structure.
$f(R)$ gravity is formally equivalent to a scalar-tensor theory where the additional degree of freedom is described by the \emph{scalaron} field $f_R \equiv \rmd f/\rmd R$~\cite{starobinsky:79, starobinsky:80}.
We parametrize our models by the background value of the scalaron field today, $\absfR$.
The $f_R$ field is massive, and below its Compton wavelength, it enhances gravitational forces by a factor of $4/3$.
Due to the density dependence of the scalaron's mass, however, viable $f(R)$ gravity models experience a mechanism dubbed the \emph{chameleon} effect~\cite{khoury:03, navarro:06, faulkner:06}, which returns gravitational forces to the standard relations in high-density regions, making them compatible with solar-system tests~\cite{hu:07a, brax:08} at $r\lesssim 20~{\rm AU}$.

The enhanced gravitational coupling can be utilized to place constraints on the $f(R)$ modification.
The transition required to interpolate between the low curvature of the large-scale structure and the high curvature of the galactic halo sets the currently strongest bound on the background field, $\absfR<|\Psi|\sim (10^{-6} - 10^{-5})$~\cite{hu:07a}, i.e., the typical depth of cosmological potential wells.
A bound of the same order is obtained from galaxies serving as strong gravitational lenses~\cite{smith:09t} at $r \sim (1 - 10)~\textrm{kpc}$
and from the comparison of nearby distances inferred from cepheids and tip of the red giant branch stars in a sample of unscreened dwarf galaxies~\cite{jain:12}.
Independently, strong constraints can also be inferred from the large-scale structure $(r\gtrsim1~\textrm{Mpc})$.
An upper bound of $\absfR\lesssim10^{-3}$, for instance, can be obtained from the cluster density profiles constrained by weak lensing measurements~\cite{lombriser:11b}.
The currently strongest constraints on $f(R)$ gravity models from the large-scale structure are inferred from the analysis of the abundance of clusters, yielding a constraint of $\absfR\lesssim10^{-4}$~\cite{schmidt:09, lombriser:10}.

It is important to note that these cluster-scale constraints have been derived by relying on a \emph{linearized} approach of the $f(R)$ modifications, i.e., assuming a linear relation between the curvature fluctuation $\delta R$ and the field fluctuation $\delta f_R$ that is correctly described by the background Compton wavelength parameter.
This approach, however, breaks down for $\absfR \lesssim 10^{-5}$, where cluster scales are affected by the chameleon mechanism.
It is therefore important for comparison to future measurements to describe the observable quantities encompassing the chameleon effect (see, e.g.~\cite{li:11b}).

Dark matter $N$-body simulations of $f(R)$ gravity provide a great laboratory for the study of the chameleon mechanism, and many efforts have been made in performing such simulations. For example, Oyaizu~\emph{et al.}~\cite{oyaizu:08a,oyaizu:08b} performed $N$-body simulations of the Hu-Sawicki~\cite{hu:07a} $f(R)$ gravity model for the first time using a particle-mesh code. Later Zhao~\emph{et al.}~\cite{zhao:10b} and Li~\emph{et al.}~\cite{li:11} simulated the same model using an adaptive particle-mesh code and significantly improved the resolution.

In this paper, we aim at finding simple analytic and semi-analytic descriptions for cluster characteristics produced in $f(R)$ $N$-body simulations in both the linearized and chameleon scenarios.
The relations we find here incorporate the chameleon mechanism and can be used to assist in the consistent comparison of $f(R)$ gravity to observations.
The outline of the paper is as follows.
In~\textsection\ref{sec:fRgravity}, we review $f(R)$ gravity with a particular focus on the Hu-Sawicki model.
In~\textsection\ref{sec:clusterproperties}, we provide (semi-)analytic relations for the scalar field, the gravitational potential, and the velocity dispersion at the virialized scales of clusters produced in linearized and chameleon $f(R)$ gravity.
Thereby, we start from the assumption of a Navarro-Frenk-White (NFW) profile for the cluster density and a power-law pseudo-phase-space density (PPSD).
\textsection\ref{sec:comparisontosimulations} is devoted to the comparison of these relations to the output of $f(R)$ gravity $N$-body simulations of gravitationally interacting cold dark matter particles.
The fit of the cluster quantities is done using three degrees of freedom, i.e., the characteristic amplitude and scale of the NFW profile, and the velocity dispersion at the characteristic scale.
We compare the values of these fitting parameters to predictions from scaling relations based on the spherical collapse and an estimation of the amplitude of the velocity dispersion employing the Jeans equation.
We discuss our results in~\textsection\ref{sec:conclusions}.
In the appendix, we give further details on the radial dependence of the PPSD profile used for the derivation of the velocity dispersion based on the expectations from the self-similar infall of a collisional gas in the context of enhanced gravitational forces such as present in $f(R)$ gravity.
We also motivate the applicability of the NFW and PPSD profile for clusters produced with modified gravitational forces.

\section{$f(R)$ gravity} \label{sec:fRgravity}

In $f(R)$ gravity, the Einstein-Hilbert action is supplemented by a free nonlinear function of the Ricci scalar $R$,
\begin{equation}
S = \frac{1}{2\kappa^2} \int \rmd^4 x \sqrt{-g} \left[ R + f(R) \right] + S_{\rm m}\left(\psi_{\rm m}; g_{\mu\nu}\right).
 \label{eq:jordan}
\end{equation}
Here, $\kappa^2 \equiv 8\pi \, G$, $\mathcal{S}_{\rm m}$ is the matter action with matter fields $\psi_{\rm m}$, and we have set $c\equiv1$. Variation with respect to the metric $g_{\mu\nu}$ yields the modified Einstein equations for metric $f(R)$ gravity,
\begin{equation}
G_{\mu\nu} + f_R R_{\mu\nu} - \left( \frac{f}{2} - \Box f_R \right) g_{\mu\nu} - \nabla_{\mu} \nabla_{\nu} f_R = \kappa^2 \, T_{\mu\nu},
\end{equation}
where the connection is of Levi-Civita type and $f_R \equiv \rmd f/\rmd R$ is the additional scalar degree of freedom of the model, characterizing the force modifications.

By a conformal transformation of the metric, the Jordan frame action Eq.~(\ref{eq:jordan}) can be recast in the Einstein frame,
\bqa
 S & = & \int \rmd^4x \sqrt{-\tg} \left[ \frac{\tilde{R}}{2\kappa^2} - \frac{1}{2} \tilde{\partial}^{\mu} \phi \tilde{\partial}_{\mu} \phi - V(\phi) \right] \nonumber \\
 & & + \Sm[\psi_{\rm m}; A^2(\phi) \tg_{\mu\nu}],
\eqa
where
\bqa
 \tg_{\mu\nu} & \equiv & (1+f_R) g_{\mu\nu}, \\
 \left( \frac{\rmd\phi}{\rmd f_R} \right)^2 & \equiv & \frac{3}{2\kappa^2} \frac{1}{(1+f_R)^2}, \label{eq:field} \\
 A(\phi) & \equiv & \frac{1}{\sqrt{1+f_R}}, \\
 V(\phi) & \equiv & \frac{f_R R - f(R)}{2\kappa^2(1+f_R)^2}. \label{eq:potential}
\eqa
Integration of Eq.~(\ref{eq:field}) gives the scalar field
\bq
 \phi = \frac{1}{\kappa} \sqrt{\frac{3}{2}} \ln (1+f_R) + \phi_0,
\eq
where we set $\phi_0\equiv0$.
Variation of the action with respect to $\phi$ yields
\bq
 \tilde{\Box} \phi = - \alpha \, \tilde{T} + V'(\phi) \equiv V'_{\rm eff}(\phi),
 \label{eq:scalardalambert}
\eq
where $\alpha = \rmd \ln A / \rmd \phi$ and $V_{\rm eff}$ is an effective potential governing the dynamics of $\phi$.
Note that $\tilde{T} = A(\phi)^4 T$.
In the quasistatic limit, we neglect time derivatives in Eq.~(\ref{eq:scalardalambert}) and we obtain the scalar field equation of interest here,
\bq
 \nabla^2 \phi = \alpha \, A(\phi)^4 \rhom + V'(\phi) \label{eq:scalarfield},
\eq
where we assumed matter dominance and use physical coordinates.

\subsection*{Hu-Sawicki model}

We specialize our considerations to the functional form of $f(R)$ proposed by Hu \& Sawicki~\cite{hu:07a},
\begin{equation}
f(R) = -\bar{m}^2 \frac{c_1 \left( R/\bar{m}^2 \right)^n}{c_2 \left( R/\bar{m}^2 \right)^n + 1},
\label{eq:husawicki}
\end{equation}
where $\bar{m}^2 \equiv \kappa^2 \, \bar{\rho}_{{\rm m}0} / 3$ and overbars refer to background quantities. The free parameters of the model $c_1$, $c_2$, and $n$ can be chosen to reproduce the $\Lambda$CDM expansion history and satisfy solar-system tests~\cite{hu:07a} through the chameleon mechanism~\cite{khoury:03, navarro:06, faulkner:06}. In the high-curvature regime, $c_2^{1/n} R \gg \bar{m}^2$, Eq.~(\ref{eq:husawicki}) simplifies to
\begin{equation}
f(R) = -\frac{c_1}{c_2} {\bar m}^2 - \frac{f_{R0}}{n} \frac{\bar{R}_0^{n+1}}{R^n},
\label{eq:backgroundmimick}
\end{equation}
where $\bar{R}_0$ denotes the background curvature today, $\bar{R}_0 = \bar{R}|_{z=0}$ , and $f_{R0} \equiv f_R(\bar{R}_0)$. We further infer
\begin{equation}
 \frac{c_1}{c_2} \bar{m}^2 = 2\kappa^2 \, \bar{\rho}_{\Lambda}
\end{equation}
from requiring equivalence with $\Lambda$CDM when $\absfR \rightarrow 0$.
From Eq.~(\ref{eq:potential}), we get
\bqa
 V(\phi) & = & \frac{R \, f_R (1+1/n) + 2 \kappa^2 \bar{\rho}_{\Lambda}}{2 \kappa^2 (1+f_R)^2}, \\
 V'(\phi) & = & \frac{ R \left( 1 - \frac{n+2}{n} f_R \right) - 4 \kappa^2 \bar{\rho}_{\Lambda}}{\sqrt{6} \kappa (1+f_R)^2},
\eqa
where $R/\bar{R}_0 = (f_{R0}/f_R)^{1/(n+1)}$.
For $f_R \ll 1$ and subtracting the background, Eq.~(\ref{eq:scalardalambert}) becomes
\bq
\nabla^2 \delta f_R = \frac{1}{3} \left[ \delta R (f_R) - \kappa^2 \, \delta \rho_{\rm m} \right], \label{eq:fR}
\eq
where $\delta f_R = f_R(R) - f_R(\bar{R})$, $\delta R = R - \bar{R}$, $\delta \rho_{\rm m} = \rho_{\rm m} - \bar{\rho}_{\rm m}$.

We assume a spatially homogeneous and isotropic cosmological background metric and consider perturbations of the Friedmann-Lema\^itre-Robertson-Walker line element in longitudinal gauge, i.e.,
$\Psi=\delta g_{00}/(2g_{00})$ and $\Phi=\delta g_{ii}/(2g_{ii})$.
Combining the time-time and time-space component of the perturbed modified Einstein equations, one obtains
\begin{widetext}
\bqa
 -3 \dot{f_R} \dot{\Psi} + 2 \nabla^2 [ (1+f_R) \Psi ] - 3 H \dot{f_R} \Psi & = & \kappa^2 (\delta \rho + 3 H v) - \left[ 6 H^2 + \frac{3\dot{f_R}^2}{(1+f_R)^2} + \nabla^2 \right] \delta f_R + \frac{\delta ( f_R R - f)}{2} + \frac{3\dot{f_R} }{1+f_R} \dot{\delta f_R}. \nonumber \\
\eqa
\end{widetext}
We assume matter dominance, i.e., $\delta\rho = \drhom$ and $v=v_{\rm m}$, the matter peculiar velocity potential defined by $\partial_{\mu} v=-\delta u_{\mu}$, where $u_{\mu}$ is the unit four-velocity.
Here, dots denote derivatives with respect to cosmic time.
For $f_R \ll 1$ and in the quasistatic limit, $\left|\nabla^2\delta f_R\right| \gg H^2 \delta f_R$, neglecting time-derivatives of the perturbations and $f_R$, this yields the modified Poisson equation
\bq
 \nabla^2 \Psi = \frac{2 \kappa^2}{3} \delta\rho_{\rm m} - \frac{1}{6} \delta R (f_R), \label{eq:pot}
\eq
where we have used $\delta f \approx f_R \delta R$ and Eq.~(\ref{eq:fR}) to replace $\nabla^2 \delta f_R$.

Note that if the background field $|f_{R0}|$ is large compared to typical gravitational potentials ($\sim 10^{-5}$), we may linearize the field equations, Eqs.~(\ref{eq:fR}) and (\ref{eq:pot}), via the approximation
\begin{equation}
\delta R \approx \left. \frac{\partial R}{\partial f_R} \right|_{R=\bar{R}} \delta f_{R} = 3 m^2 \delta f_R, \label{eq:linapp}
\end{equation}
where $m$ is the mass of the $f_R$ field at the background.
In Fourier space, the solution to Eqs.~(\ref{eq:fR}) and (\ref{eq:pot}) within the linearized approximation is
\begin{equation}
k^2 \Psi({\bf k}) = -\frac{\kappa^2}{2} \left\{ \frac{4}{3} - \frac{1}{3} \left[ \left( \frac{k}{m\,a} \right)^2 + 1 \right]^{-1} \right\} a^2 \delta \rhom({\bf k}),
\label{eq:linearized}
\end{equation}
for a comoving wavenumber $k=|{\bf k}|$.
For scales $k \gg m \, a$, this leads to an enhancement of gravitational forces by a factor of $4/3$. Computations using Eq.~(\ref{eq:linearized}) are referred to as the \emph{no-chameleon} or \emph{linearized} $f(R)$ case~\cite{oyaizu:08b}, whereas solutions to Eqs.~(\ref{eq:fR}) and (\ref{eq:pot}) are referred to as \emph{chameleon} $f(R)$ gravity.
In the following section, we will study solutions for the scalar field $\delta f_R$ and the Newtonian potential $\Psi$ within a virialized cluster in both scenarios.

\section{Cluster quantities} \label{sec:clusterproperties}

Effects from $f(R)$ modifications of gravity on halo properties were studied in, e.g.,~\cite{martino:08, schmidt:08, schmidt:10, borisov:11, li:12, li:11b}.
The enhanced abundance of clusters caused by the modification was used in~\cite{schmidt:09, lombriser:10} to place an upper bound on the scalaron background value of $\absfR\lesssim10^{-4}$.
Ref.~\cite{lombriser:11b} used cluster-galaxy lensing measurements of the excess surface mass density to constrain $\absfR$ based on the $f(R)$ enhancement of the cluster density profile around the virial radius (see~\cite{schmidt:08}), finding an upper bound of $\absfR\lesssim10^{-3}$.
These analyses have been carried out in the linearized regime of $f(R)$ gravity, i.e., where the approximation Eq.~(\ref{eq:linapp}) is valid.
However, with future measurements, constraints will penetrate into the chameleon regime and it becomes important to incorporate the effects of the chameleon mechanism on the observables.
This shall be the concern of this section.
Based on the NFW profile, we derive here analytic formulae, e.g., for the gravitational potential and velocity dispersion as observed within the virialized cluster.
These relations can subsequently be used to predict observables in $f(R)$ gravity without having to rely directly on simulations and constrain $\absfR$ in the chameleon regime without the necessity of running a large number of simulations.

\subsection{Density}

Navarro, Frenk, and White (NFW)~\cite{navarro:97} found that the simple relation
\bq
 \drhom(r) = \frac{\beta \, \rhoc}{\frac{r}{\rs} \left( 1 + \frac{r}{\rs} \right)^2},
 \label{eq:nfw}
\eq
provides good fits to the cluster density profiles measured in Newtonian CDM simulations.
Here, $\rs$ denotes the characteristic scale and $\rhos=\beta \, \rhoc$ is the characteristic density with the critical background density $\rhoc$.
As we will show in~\textsection\ref{sec:comparisontosimulations}, this simple function provides comparably good fits to $f(R)$ gravity simulations in both the linearized and chameleon scenarios for $r\in(r_0,\rvir)$, where $\rvir$ is the virial radius of the cluster and $r_0$ is conservatively set by the requirement that $N>800$ particles in the simulations fall within that radius (see~\textsection\ref{sec:comparisontosimulations} and Appendix~\ref{sec:scaling}).
In Appendix~\ref{sec:selfsiminfall}, we argue that the applicability of the NFW profile to modified gravity can be motivated by consideration of the self-similar infall of a collisional gas under modified forces, preserving consistency with the Jeans equation and the virial theorem.

In the following, we shall assume this profile to be exact on the scales of interest, $r\in(r_0,\rvir)$, and derive from that (semi-)analytic relations for the mass, the scalar field, the gravitational potential, and the velocity dispersion.

\subsection{Mass}

We integrate Eq.~(\ref{eq:nfw}) from the origin of the cluster to the radius $r$ to obtain the mass of the cluster enclosed by $r' \leq r$.
However, the NFW profile might not apply for $r' \leq r_0$. For fair comparison with simulation results, we add a correction for the inner part of the cluster.
The mass can then be obtained by the integration
\bqa
 M(r) & = & 4 \pi \int_0^r \rmd r' \, r'^2 \left[ \drhom(r') + \Delta \rhom(r') \Theta(r_0-r') \right] \nonumber \\
 & = & 4 \pi \, \beta \, \rhoc \rs^3 \left[ \ln\left( 1 + \frac{r}{\rs} \right) - \frac{r}{r+\rs} \right] + M_{\rm c},
 \label{eq:mass}
\eqa
where $\Delta \rho_{\rm m}(r) = \delta\rho_{\rm m, sim}(r) - \delta\rho_{\rm m, NFW}(r)$ and $M_{\rm c}$ defines a mass calibration at $r_0$.
Eq.~(\ref{eq:mass}) applies to Newtonian as well as $f(R)$ gravity.

\subsection{Scalar field} \label{sec:scalarfield}

\begin{figure*}
 \resizebox{\hsize}{!}{
  \includegraphics{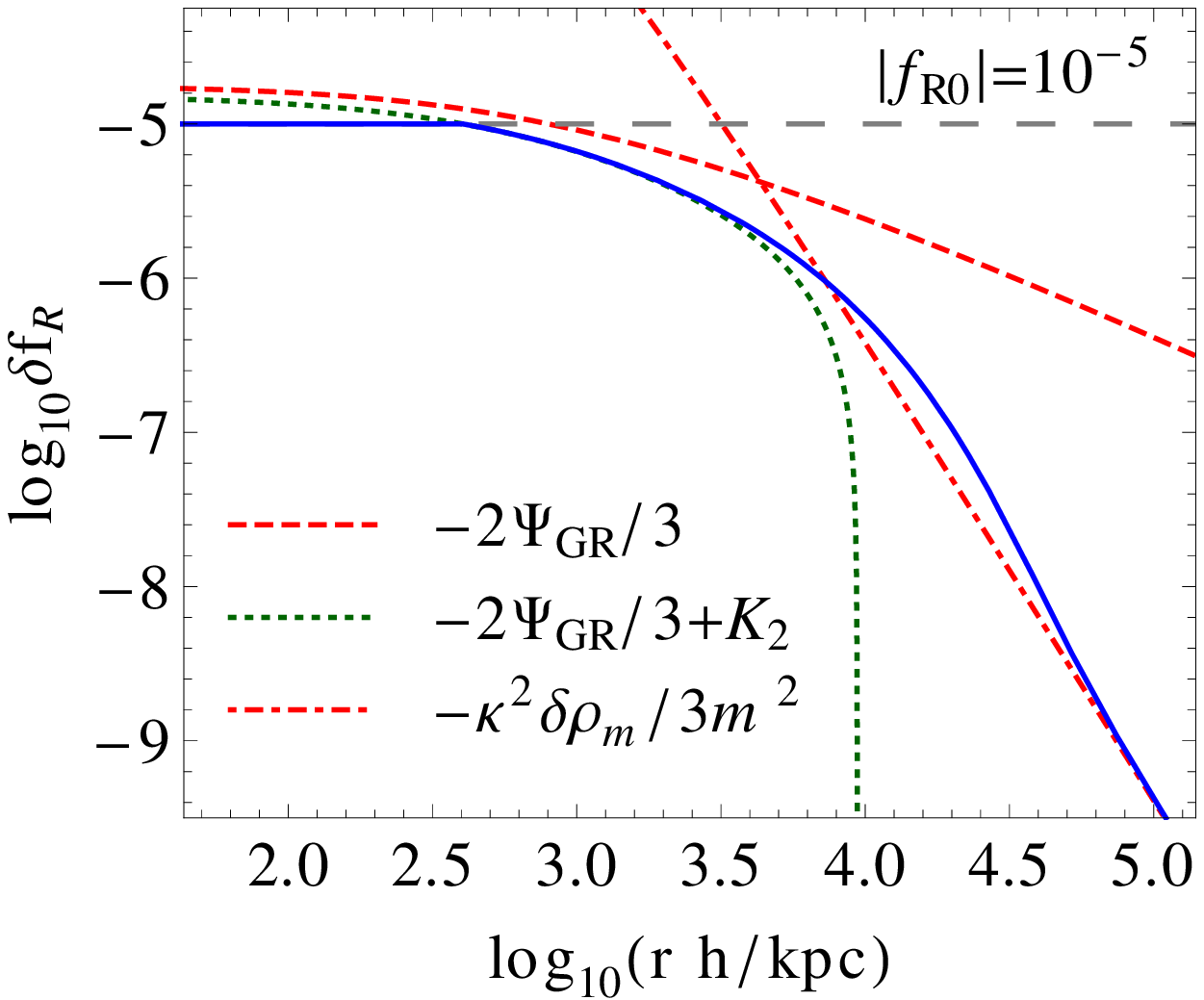}
  \resizebox{0.735\hsize}{!}{\includegraphics{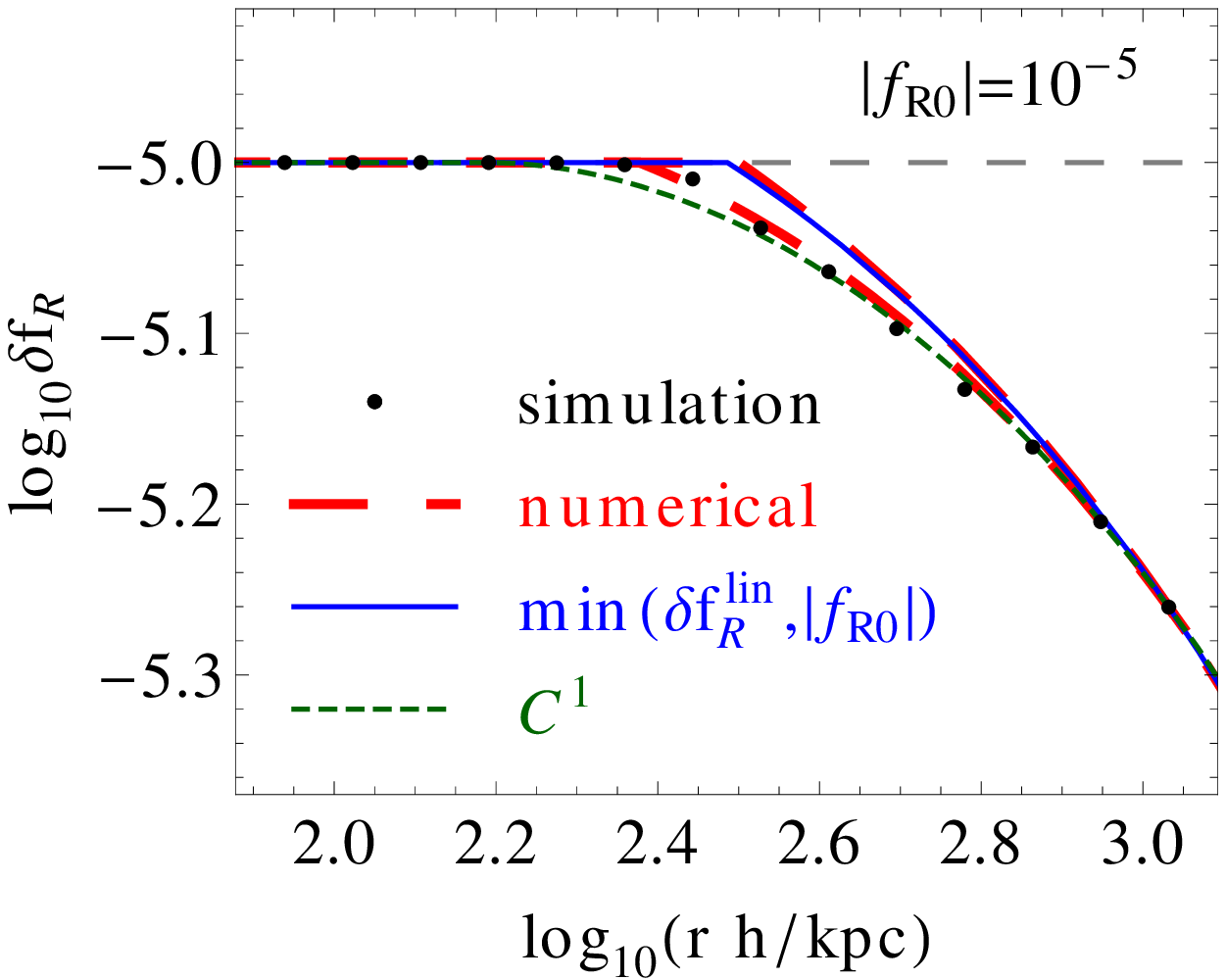}}
  \includegraphics{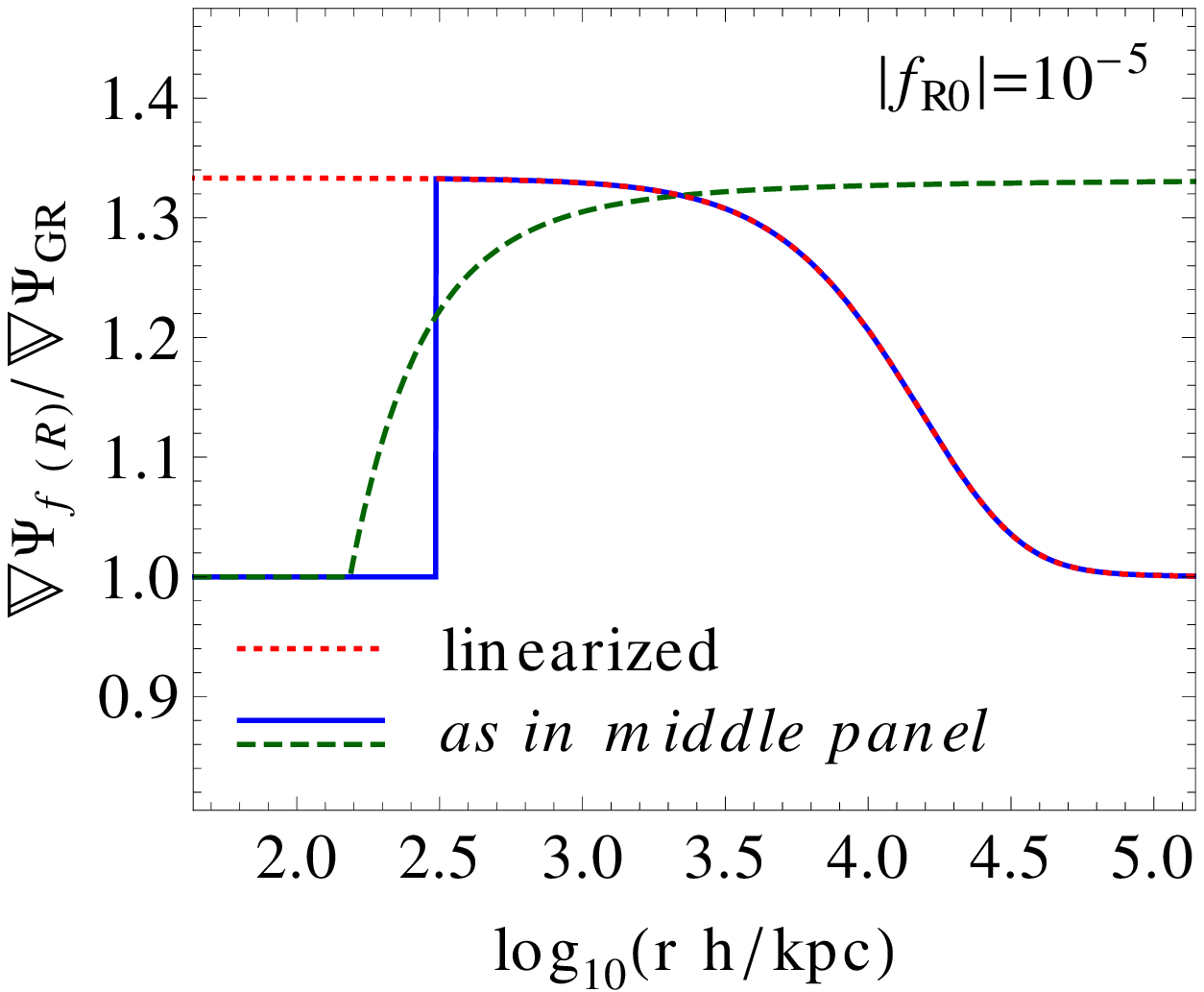}
 }
\caption{
Properties of the scalar field $\dfR$ for a cluster of $M_{\rm vir}=1.36\times10^{14}~\Msunh$ and a background field amplitude of $\absfR=10^{-5}$ (long-dashed line) with $n=1$.
\emph{Left}: The analytically derived scalar field $\delta f_R$ (solid line) and its behavior at the limit of large and small scales, respectively.
The transition of the linearized field into a chameleon field occurs instantaneously where $\dfR = \absfR$.
\emph{Middle}: Comparison between the instantaneous chameleon transition (solid line) and the $\mathcal{C}^1$-transition for $\dfR$ (dashed line) with analog semi-analytical derivation to~\cite{pourhasan:11}.
Matching $\dfR = \delta f_{R,{\rm sim}}$ at $\rvir$ brings the approximations into good agreement with the simulated $\dfR(r)$ (dots). The thick long-dashed lines show numerical solutions to Eq.~(\ref{eq:fR}) assuming a NFW profile with integration constraints set by additionally requiring that $\dfR'(\rvir)$ corresponds to the analytic and semi-analytic result, respectively.
\emph{Right}: Enhanced force by the linearized (dotted line), the instantaneous chameleon (solid line), and the $\mathcal{C}^1$ (dashed line) chameleon scalar field  $\delta f_R$ (matched at $\rvir$), respectively.
The lines of the linearized and instantaneous chameleon field overlap beyond the chameleon region.
}
\label{fig:scalaron}
\end{figure*}

We first derive the scalar field $\dfR$ in the linearized case and based on this result, we then construct an analytic approximation for the solution in the chameleon case by requiring that $\dfR \ngtr \absfR$.
For comparison, we also study the approximation of the chameleon scalar field of Pourhasan \emph{et al.}~\cite{pourhasan:11} and a numerical solution to the scalar field equation, Eq.~(\ref{eq:fR}), assuming spherical symmetry and the applicability of the NFW halo density profile.
For clarity, we shall denote our solutions for the scalar field as $\dfRlin$ and $\dfRcham$ for the linearized and chameleon case, respectively.

\subsubsection{Linearized field}

In order to find $\dfRlin$, we solve Eq.~(\ref{eq:fR}) with the approximation Eq.~(\ref{eq:linapp}) and the assumption that $\drhom$ is described by a NFW profile.
Furthermore assuming spherical symmetry, we obtain the differential equation
\bq
 \left( \partial_r^2 + \frac{2}{r} \partial_r \right) \dfRlin - m^2 \dfRlin + \frac{\kappa^2}{3} \frac{\rhos \, \rs^3}{r(r+\rs)^2} = 0,
 \label{eq:fRlindiff}
\eq
which has the solution
\begin{widetext}
\bq
 \dfRlin(r) = - \left[ \frac{\kappa^2}{6} \rhos r_s^3 \left\{ \Gamma\left[0,m(r+\rs)\right] e^{2 m (r+\rs)} + \Gamma\left[0,-m(r+\rs)\right] \right\}
 + \left( C_1 + \frac{C_2}{2m} e^{2 m \, r} \right) e^{m \, \rs} \right] \frac{e^{-m(r+\rs)}}{r}.
 \label{eq:fRlin}
\eq
\end{widetext}
Here, $C_1$ and $C_2$ are integration constants and
\bq
 \Gamma(s,r) = \int_{r}^{\infty} \rmd t \, t^{s-1} e^{-t}
\eq
denotes the upper incomplete gamma function.
$C_2$ is the amplitude of a growing mode and since we want to restore general relativity (GR) at $r\rightarrow\infty$, we set $C_2=0$.
We further require $\kappa^2 \drhom/3$ to dominate over $m^2 \dfRlin$ towards the origin of the halo and hence,
\bq
 \lim_{r \rightarrow 0} r \, \dfRlin \rightarrow 0. \label{eq:dfRorigin}
\eq
This is also apparent from the resemblance of Eq.~(\ref{eq:fRlindiff}) when $m^2 \dfRlin$ is small to the standard Poisson equation and its solution for the Newtonian potential $\PsiGR$ (see~\textsection\ref{sec:gravpot}).
The integration constant then becomes
\bq
 C_1 = - \frac{\kappa^2 \rhos r_{\rm s}^3}{6} e^{-m \, \rs} \left[ e^{2m\,\rs} \Gamma(0,m\,\rs) + \Gamma(0,-m\,\rs) \right]
\eq
and hence, we arrive at our solution for the linearized scalar field
\bqa
 \dfRlin(r) & = & -\frac{\kappa^2 \rhos r_s^3}{6} \left\{ \Gamma[0,m(r+\rs)] e^{2 m (r+\rs)} \right. \nonumber \\
 & & + \Gamma[0,-m(r+\rs)] - \Gamma(0,-m \, \rs) \nonumber \\
 & & \left. -e^{2 m \, \rs} \Gamma(0,m \, \rs) \right\} \frac{e^{-m(r+\rs)}}{r}. \label{eq:dfRlin}
\eqa
Note that at scales where $\rhom \sim \brhom$ and $\rhom \gg \brhom$, we obtain the limits
\bqa
 \dfRlin & = & \frac{\kappa^2}{3m^2} \drhom, \label{eq:dfRlinlimlarge} \\
 \nabla^2 \dfRlin & = & -\frac{\kappa^2}{3} \drhom, \label{eq:dfRlinlimsmall}
\eqa
respectively.
Eq.~(\ref{eq:dfRlinlimsmall}) implies that
\bq
 \dfRlin = \frac{\kappa^2 \rhos \rs^3}{3} \frac{\ln(1+r/\rs)}{r} + \frac{K_1}{r} + K_2,
 \label{eq:exterior}
\eq
where the integration constants $K_1=0$ due to Eq.~(\ref{eq:dfRorigin}) and
$K_2 = - m \, \kappa^2 \rhos \rs^3 \exp(m \, \rs) \Gamma(0,m\,\rs)/3$
to match Eq.~(\ref{eq:dfRlin}) at the origin.
Hence, for $\rhom \gg \brhom$,
\bq
 \dfRlin = -\frac{2}{3} \PsiGR - \frac{\kappa^2 \rhos \rs^3 }{3} m \, e^{m \, r_{\rm s}} \Gamma(0,m\,r_{\rm s}),
\eq
where $\PsiGR$ is taken for an isolated halo assuming a NFW profile on all scales (see~\textsection\ref{sec:gravpot}).

We illustrate our solution for the scalar field and its behavior at the limit of large and small scales, respectively, in the left panel of Fig.~\ref{fig:scalaron}.

\subsubsection{Chameleon field}

In order to describe the chameleon mechanism in $f(R)$ gravity, let us revisit Eq.~(\ref{eq:scalarfield}), which for $|f_R| \ll 1$ and $\alpha = -\kappa/\sqrt{6}$ becomes
\bq
 \nabla^2 \phi = -\frac{\kappa}{\sqrt{6}} \rhom + V'(\phi).
 \label{eq:chamscal}
\eq
In high density regions, where $\kappa \, \rhom \gg -\sqrt{6} \, \nabla^2 \phi$, the scalar field becomes
\bq
 \sqrt{\frac{2}{3}}\kappa\,\phi = f_{R0} \left[ \frac{\bar{R}_0}{\kappa^2(\rhom+4\bar{\rho}_{\Lambda})} \right]^{n+1} = f_R.
 \label{eq:interior}
\eq
Since $\rhom \gg \rhoc$, for $n>-1$, we get $f_R \sim \kappa \, \phi \simeq 0$.
Hence, modifications of gravity are suppressed.

We consider three different approaches for describing the transition of $\dfRlin$ to $\dfRcham$ and compare them with each other.
The first approach is the assumption of a simplified, but analytically describable, instantaneous transition to $\dfR=-f_{R0}$, whereas the second is a semi-analytical match of the two regions and the third is a numerical solution to Eq.~(\ref{eq:fR}).
Thus, in the first case, in order to implement the chameleon mechanism in our fit, we may simply require
\bq
 \fRcham = \min \left( \fRlin, 0 \right)
 \label{eq:fRcham}
\eq
or equivalently,
$\dfRcham = \min \left( \dfRlin, -f_{R0} \right)$ (see left and middle panels of Fig.~\ref{fig:scalaron}).
As we will show in~\textsection\ref{sec:comparisontosimulations}, this yields a good approximation to the simulated chameleon $f_R$ field.
The difference being a more gradual decrease in $\partial_r f_R$ from $-2\partial_r \PsiGR/3$ to $0$ instead of an instantaneous transition (cf. right panel of Fig.~\ref{fig:scalaron}).
The chameleon transition is, however, very efficient, which allows the applicability of this approach.

For comparison, we follow the semi-analytic approach of Pourhasan \emph{et al.}~\cite{pourhasan:11} for describing the chameleon field for an inverse power-law potential of a scalar field. Their procedure corresponds to matching the $f(R)$ chameleon interior solution, Eq.~(\ref{eq:interior}), applying to $r\in(r_-,r_+)$ to the chameleon exterior solution, Eq.~(\ref{eq:exterior}), for $r > r_{\rm c}$ at the transition scale $r_{\rm c}$.
This defines the integration constants in Eq.~(\ref{eq:exterior}).
$K_2=0$ follows from the requirement that $\delta f_R \rightarrow 0$ when $r \rightarrow \infty$.
$K_1$ and $r_{\rm c}$ are determined by requiring that
\bq
 \dfR(r) \equiv \left( \delta f_{R,r<r_{\rm c}}^{\rm in} \cup \delta f_{R,r \geq r_{\rm c}}^{\rm out} \right)(r) \in \mathcal{C}^1(U)
 \label{eq:dfRC1}
\eq
with $r_{\rm c} \in U \subset \mathbb{R}^+_0$, i.e., the matched scalar field and its derivative are continuous at the transition.
We compute $r_{\rm c}$ numerically and show the according solution for $\dfR$ in the middle panel of Fig.~\ref{fig:scalaron}.
Note that Eq.~(\ref{eq:dfRC1}) assumes that the interior solution also applies to the shell $r \in (r_+, r_{\rm c})$, where $r_+ < r_{\rm c}$.
Following~\cite{pourhasan:11}, the boundaries of the interior region, i.e., the regime of applicability of Eq.~(\ref{eq:interior}),  $r \in (r_-, r_+)$, can be obtained from the roots of $-3 \nabla^2 f_R / \kappa^2 \rhom = 10^{-2}$. If for $r_+<r_{\rm c}$, we have $(r_{\rm c} - r_+) \ll r_{\rm c}$, the interior solution may be extended into the shell, which in this case is sufficiently thin.
Strictly speaking, the condition $r_+ > r_{\rm c}$ is not satisfied for the scalar field shown in Fig.~\ref{fig:scalaron}.
The $\dfR$ computed with this procedure, however, still yields a good description to the simulated scalar field (see middle panel of Fig.~\ref{fig:scalaron}).

As a third case, we numerically solve the differential equation for $\dfR$, Eq.~(\ref{eq:fR}),
assuming spherical symmetry and that $\drhom$ is given by a NFW profile.
For stability reasons we use the substitution $f_R = - e^{u(r)}$ (see~\cite{zhao:10b}) in our computations.
We compare the numerical solutions for $\dfR$ obtained in this way to the chameleon scalar fields obtained through the analytic and semi-analytic approach, Eq.~(\ref{eq:fRcham}) and Eq.~(\ref{eq:dfRC1}), respectively, in the middle panel of Fig.~\ref{fig:scalaron}.

Note that due to the limited applicability of the NFW density profile for the description of $\drhom$ beyond $r \in (r_0,\rvir)$ and the unknown environment, when comparing to simulations, we correct the analytic and semi-analytic solutions of $f_R$ to match them to $f_{R, \rm sim}$ at $\rvir$, or equivalently, we require this constraint when numerically solving Eq.~(\ref{eq:fR}).
In specific, this means adding a constant deviation from the background density in Eq.~(\ref{eq:fRlindiff}) and dropping the condition $K_2=0$ in the instantaneous and $\mathcal{C}^1$ chameleon solution, respectively.
This brings the computed scalar fields into good agreement with the simulated $\dfR$ over the radial range $r \in (r_0,\rvir)$ as is demonstrated in the middle panel of Fig.~\ref{fig:scalaron}.
For the comparison to the numerical solution of Eq.~(\ref{eq:fR}), we further constrain the integration by requiring that $\dfR'(\rvir)$ corresponds to the derivative of the solution
for the $\mathcal{C}^1$ and the instantaneous chameleon field, i.e., the semi-analytic and analytic $\dfR$, respectively.

\begin{figure}
 \resizebox{0.75\hsize}{!}{
  \includegraphics{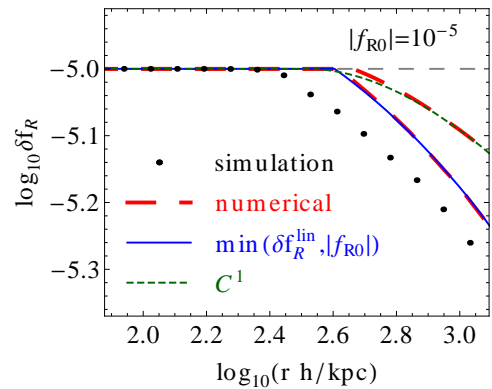}
 }
\caption{Same as middle panel of Fig.~\ref{fig:scalaron} but with cosmological background density as boundary condition instead of matching to $\delta f_{R, {\rm sim}}(\rvir)$ for $\dfR$. For the radial derivatives, we further require that $\dfR'(\rvir)$ corresponds to the analytic and semi-analytic result, respectively.
}
\label{fig:scalaron2}
\end{figure}

It is important to note that matching $\dfR(\rvir)$ to simulations is essential for recovering the radial dependence of the scalar field from simulations.
As demonstrated in Fig.~\ref{fig:scalaron2}, if we use the cosmological background as the boundary condition instead, i.e., $K_2=0$ in Eq.~(\ref{eq:exterior}), we are not able to reproduce the scalar field $\delta f_{R,{\rm sim}}$.

\subsection{Gravitational potential} \label{sec:gravpot}

For a spherically symmetric mass distribution, the gravitational potential at $r \in (r_0,r_{\rm vir})$ in Newtonian gravity is obtained by the sum of the interior and exterior spherical mass shells, i.e.,
\bqa
 \Psi_{\rm GR} & = & - \frac{\kappa^2}{8 \pi} \left[ \frac{1}{r} \int_0^r \rmd M(r') - \int_r^{\infty} \frac{\rmd M(r')}{r'} \right] \\
 & = & -\frac{\kappa^2}{2} \left\{ \frac{1}{r} \int_0^r \rmd r' \, r'^2 \left[ \rhom(r') + \Delta \rhom(r') \theta(r_0-r') \right] \right. \nonumber \\
 & & \left. + \int_r^{\infty} \rmd r' \, r' \left[ \rhom(r') + \Delta \rhom(r') \theta(r'- \rvir) \right] \right\} \\
 & = & - \frac{\kappa^2 \rhos \rs^3}{2} \frac{\ln(1+r/\rs)}{r} - \frac{\kappa^2 M_{\rm c}}{8\pi \, r} + \Psi_{\rm c} + \Psi_{\rm ext}.
 \label{eq:GRpot}
\eqa
Here, $\Psi_{\rm ext}$ indicates an external gravitational field in case the halo is not isolated
and $\Psi_{\rm c}$ accounts for deviations from the NFW density profile at $r>\rvir$, e.g., the two-halo contribution, i.e., $\Psi_{\rm c}$ does not vanish even if the halo is isolated.

Combining Eqs.~(\ref{eq:fR}) and (\ref{eq:pot}) yields
\bqa
 \nabla^2 \Psi & = & \frac{\kappa^2}{2}\delta\rho_{\rm m} - \frac{1}{2}\nabla^2\delta f_R \\
 & = & \nabla^2 \left( \Psi_{\rm GR} - \frac{1}{2}\delta f_R \right).
\eqa
By partial integration and with the analog boundary conditions as in the integration of $\PsiGR$, i.e.,
\bqa
 \lim_{r\rightarrow0} r \, \partial_r \dfR = 0, \\
 \lim_{r\rightarrow\infty} \partial_r \dfR = 0,
\eqa
we obtain the modified gravitational potential
\bq
 \Psi = \PsiGR - \frac{1}{2} \dfR \label{eq:fRgravpot}.
\eq
For comparison to simulations, the
combination $\Psi_{\rm c}+\Psi_{\rm ext}$,
i.e., the halo density correction to the NFW profile beyond $\rvir$ and the external gravitational field $\Psi_{\rm ext}$ from the environment,
is calibrated to $\Psi_{\rm sim}(\rvir)$.

\subsection{Velocity dispersion}

From contemplations on the self-similar secondary infall and the shocked accretion of a collisional gas onto the center of an initially spherical uniform overdensity in an otherwise uniformly expanding Einstein-de Sitter universe, Bertschinger~\cite{bertschinger:85} determined power-law behaviors for the fluid variables, which produce the phase-space density profile
\bq
 \frac{\rho^{5/2}}{p^{3/2}} \propto r^{-15/8},
 \label{eq:psd}
\eq
where $p$ denotes the pressure of the gas.
We refer the reader to Appendix~\ref{sec:selfsiminfall} for details about this derivation and its applicability to modified gravity.
With a gas of pressure $p=\rhom \, \sigma^2$, where $\sigma$ is the velocity dispersion, Eq.~(\ref{eq:psd}) yields the pseudo phase-space density (PPSD) profile
\bq
 \frac{\rhom}{\sigma(r)^3} = \frac{1}{4} \frac{\rhos}{\sigmas^3} \left(\frac{\rs}{r}\right)^{15/8}
 \label{eq:ppsd}
\eq
with $\sigmas\equiv\sigma(\rs)$.
The simple relation Eq.~(\ref{eq:ppsd}) was found to give good fits to Newtonian CDM simulations~\cite{taylor:01}.

As pointed out in Appendix~\ref{sec:selfsiminfall}, the $r$-dependence of the PPSD profile does not change in a simplified approach of $f(R)$ gravity, i.e., where a simple force amplification of $4/3$ is assumed.
We therefore use the velocity dispersion
\bq
 \sigma^2(r) = \sigmas^2 \left( \frac{4\rhom}{\rhos} \right)^{2/3} \left( \frac{r}{\rs} \right)^{5/4}
 \label{eq:velocitydisp}
\eq
to compare to simulations, where $\sigmas$ is taken to be an additional degree of freedom that we fit to simulations and $\rhom$ is assumed to be correctly described by the NFW density profile for $r\in(r_0,\rvir)$.

Note that the radial effect on $\sigma^2$ from a transition in the modified forces is blurred out over a wide range of scales.
As shown in~\textsection\ref{sec:comparisontosimulations}, the radial dependence of Eq.~(\ref{eq:velocitydisp}) fits the chameleon $f(R)$, linearized $f(R)$, and $\Lambda$CDM simulations equally well.
For the estimation of $\sigmas$, however, in order to encompass the chameleon as a function of halo mass, we can replace the constant force enhancement with a weighted average over the modification shown in the right panel of Fig.~\ref{fig:scalaron} (cf.~\cite{schmidt:10, clampitt:11}).

\section{Comparison to simulations} \label{sec:comparisontosimulations}

Based on the NFW fit for the halo density profile, Eq.~(\ref{eq:nfw}), we have constructed in~\textsection\ref{sec:clusterproperties} analytic fits to the halo mass enclosed at radius $r$, Eq.~(\ref{eq:mass}), the scalar field $\dfR$, Eqs.~(\ref{eq:dfRlin}) and (\ref{eq:fRcham}), and the modified gravitational potential given by Eqs.~(\ref{eq:GRpot}) and (\ref{eq:fRgravpot}).
We have further assumed that the velocity dispersion of dark matter particles is correctly described by the power-law PPSD profile predicted by the self-similar collapse of a collisional gas.

In this section, we shall test these relations against collisionless dark matter $N$-body simulations of Newtonian, linearized $f(R)$, and chameleon $f(R)$ gravity.
Thereby, we assume the overdensity $\beta$ (or equivalently $\rhos$) and the characteristic scale $\rs$ in the NFW profile, as well as the velocity dispersion at $\rs$, i.e., $\sigmas$, to be free fitting parameters.
We then compare the quality of these fits between the different simulation outputs and analyze the ability of scaling relations based on the spherical collapse to give predictions for $\rhos$, $\rs$, and $\sigmas$.

\subsection{$N$-body simulations} \label{sec:simulations}

The simulations used in this work were carried out for the Newtonian (GR), linearized (N), and full chameleon (F) scenarios for each field strength $\absfR=10^{-6}, 10^{-5}, 10^{-4}$ with $n=1$~\cite{zhao:10b}.
Each set of simulations consists of 10 realizations with each box size, $L_{\rm box} = 64\hMpc, \ 128\hMpc, \ 256\hMpc$, and a total particle number of $N_{\rm p}=256^3$ placed on $128^3$ domain grids.
Thereafter, the different set of simulations are denoted by GR-$[L_{\rm box}]$, N-$[-\log_{10}\absfR]$-$[L_{\rm box}]$, and F-$[-\log_{10}\absfR]$-$[L_{\rm box}]$.
During the simulation, the domain grids are refined progressively where the local densities are sufficiently large to reach a predefined threshold. In this way, the grid structure efficiently follows the density distribution so that the high density regions can be well resolved.
The cosmological parameters are fixed to values following the
WMAP 3-year results,
$\Omega_{\Lambda}=0.76$, $\Omega_{\rm m}=1-\Omega_{\Lambda}$, $h=0.73$, $n_{\rm
  s}=0.958$, and the initial power in curvature fluctuations $A_{\rm
  s}=(4.89\times 10^{-5})^2$ at $k=0.05~\textrm{Mpc}^{-1}$.

Halos within the simulation and their associated masses are identified via a spherical overdensity (SO) algorithm (cf.~\cite{jenkins:00}). The particles are placed on the grid by a cloud-in-cell interpolation and counted within a growing sphere around the center of mass until the required overdensity is reached. The mass of the halo is then defined by the sum of the particle masses contained in the sphere. This process is started at the highest overdensity grid point and hierarchically continued to lower overdensity grid points until all halos are identified.
Note that the virial overdensity obtained for $\Lambda$CDM is used to identify halos even in $f(R)$ gravity in order to make a fair comparison between the different models (see Appendix~\ref{sec:scaling}).

\subsection{Performance of fits} \label{sec:results}

\begin{table*}
 \begin{tabular}{lccccccc}
  Simulation & Mass range & $\langle M \rangle$ & $\sqrt{\left< \chi^2_{\drhom, {\rm red}} \right>}$ & $\sqrt{\left< \chi^2_{M, {\rm red}} \right>}$ & $\sqrt{\left< \chi^2_{\sigma, {\rm red}} \right>}$ & $\sqrt{\left< \chi^2_{\delta f_R, {\rm red}} \right>}$ & $\sqrt{\left< \chi^2_{\Psi, {\rm red}} \right>}$ \\
  \hline
  \hline
  GR-64   & $2 \times 10^{13} - 2 \times 10^{14}$ & $4.25 \times 10^{13}$ & 10 & 17 & 10 & $\cdots$ & $\cdots$ \\
  GR-128  & $1 \times 10^{14} - 5 \times 10^{14}$ & $1.74 \times 10^{14}$ & 8  & 14 & 9  & $\cdots$ & $\cdots$ \\
  GR-256  & $3 \times 10^{14} - 7 \times 10^{14}$ & $4.14 \times 10^{14}$ & 5  & 9  & 9  & $\cdots$ & $\cdots$ \\
  \hline
  N-64-4  & $2 \times 10^{13} - 2 \times 10^{14}$ & $4.44 \times 10^{13}$ & 10 & 18 & 10 & 1 & 6 \\
  N-128-4 & $1 \times 10^{14} - 5 \times 10^{14}$ & $1.84 \times 10^{14}$ & 8  & 15 & 9  & 1 & 5 \\
  N-256-4 & $3 \times 10^{14} - 7 \times 10^{14}$ & $4.21 \times 10^{14}$ & 5  & 10 & 8  & 1 & 3 \\
  \hline
  N-64-5  & $2 \times 10^{13} - 2 \times 10^{14}$ & $4.35 \times 10^{13}$ & 10 & 17 & 9  & 4 & 6 \\
  N-128-5 & $1 \times 10^{14} - 5 \times 10^{14}$ & $1.82 \times 10^{14}$ & 8  & 15 & 9  & 4 & 5 \\
  N-256-5 & $3 \times 10^{14} - 7 \times 10^{14}$ & $4.19 \times 10^{14}$ & 5  & 10 & 9  & 4 & 4 \\
  \hline
  N-64-6  & $2 \times 10^{13} - 2 \times 10^{14}$ & $4.45 \times 10^{13}$ & 10 & 17 & 10 & 7 & 6 \\
  N-128-6 & $1 \times 10^{14} - 5 \times 10^{14}$ & $1.76 \times 10^{14}$ & 8  & 14 & 10 & 7 & 6 \\
  N-256-6 & $3 \times 10^{14} - 7 \times 10^{14}$ & $4.14 \times 10^{14}$ & 5  & 9  & 10 & 5 & 3 \\
  \hline
  F-64-4  & $2 \times 10^{13} - 2 \times 10^{14}$ & $4.50 \times 10^{13}$ & 12 & 17 & 10 & 6 & 6  \\
  F-128-4 & $1 \times 10^{14} - 5 \times 10^{14}$ & $1.81 \times 10^{14}$ & 10 & 14 & 8  & 5 & 5  \\
  F-256-4 & $3 \times 10^{14} - 7 \times 10^{14}$ & $4.18 \times 10^{14}$ & 5  & 10 & 8  & 4 & 3  \\
  \hline
  F-64-5  & $2 \times 10^{13} - 2 \times 10^{14}$ & $4.33 \times 10^{13}$ & 10 & 18 & 9  & 4 & 6  \\
  F-128-5 & $1 \times 10^{14} - 5 \times 10^{14}$ & $1.71 \times 10^{14}$ & 7  & 15 & 7  & 4 & 4  \\
  F-256-5 & $3 \times 10^{14} - 7 \times 10^{14}$ & $4.15 \times 10^{14}$ & 5  & 10 & 7  & 2 & 3  \\
  \hline
  F-64-6  & $2 \times 10^{13} - 2 \times 10^{14}$ & $4.38 \times 10^{13}$ & 10 & 16 & 8 & 1   & 5 \\
  F-128-6 & $1 \times 10^{14} - 5 \times 10^{14}$ & $1.73 \times 10^{14}$ & 8  & 14 & 8 & 0.1 & 4 \\
  F-256-6 & $3 \times 10^{14} - 7 \times 10^{14}$ & $4.12 \times 10^{14}$ & 5  & 9  & 9 & 0.1 & 4
 \end{tabular}
 \caption{Comparison of goodness of fit of the analytic predictions of~\textsection\ref{sec:clusterproperties} based on the NFW halo density profile and the power-law PPSD between the GR, linearized $f(R)$, and full chameleon $f(R)$ simulations. Cluster masses are given in $\Msunh$. $\sqrt{\chi^2_{\rm red}}$ is computed from the \%-deviation from simulations for the halo profile $\drhom$, the mass $M$, the velocity dispersion $\sigma$, the scalar field $\delta f_R$, and the gravitational potential $\Psi$ in $r\in(r_0,\rvir)$.}
 \label{tab:deviations}
\end{table*}

\begin{figure*}
 \resizebox{\hsize}{!}{\includegraphics{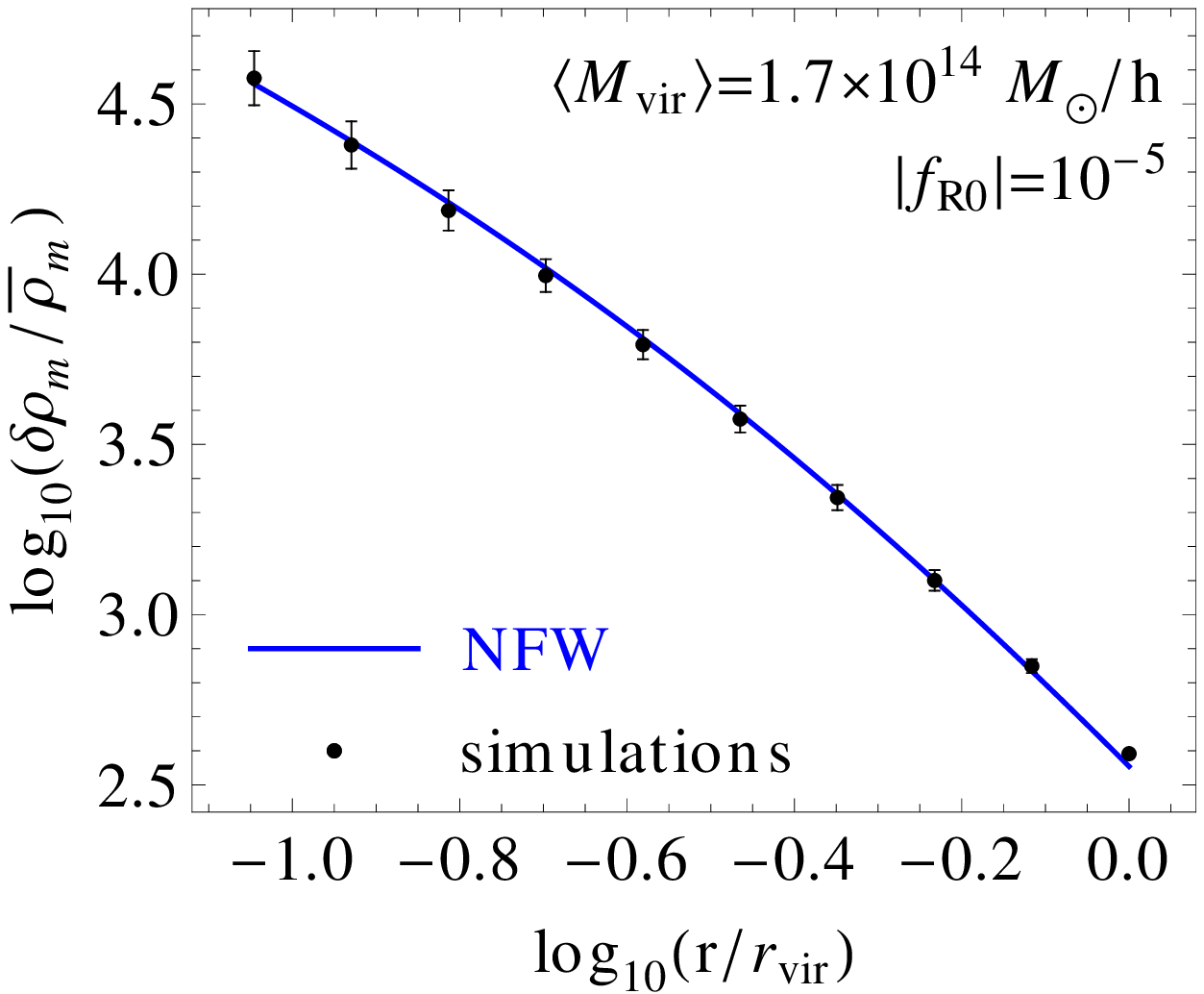}\includegraphics{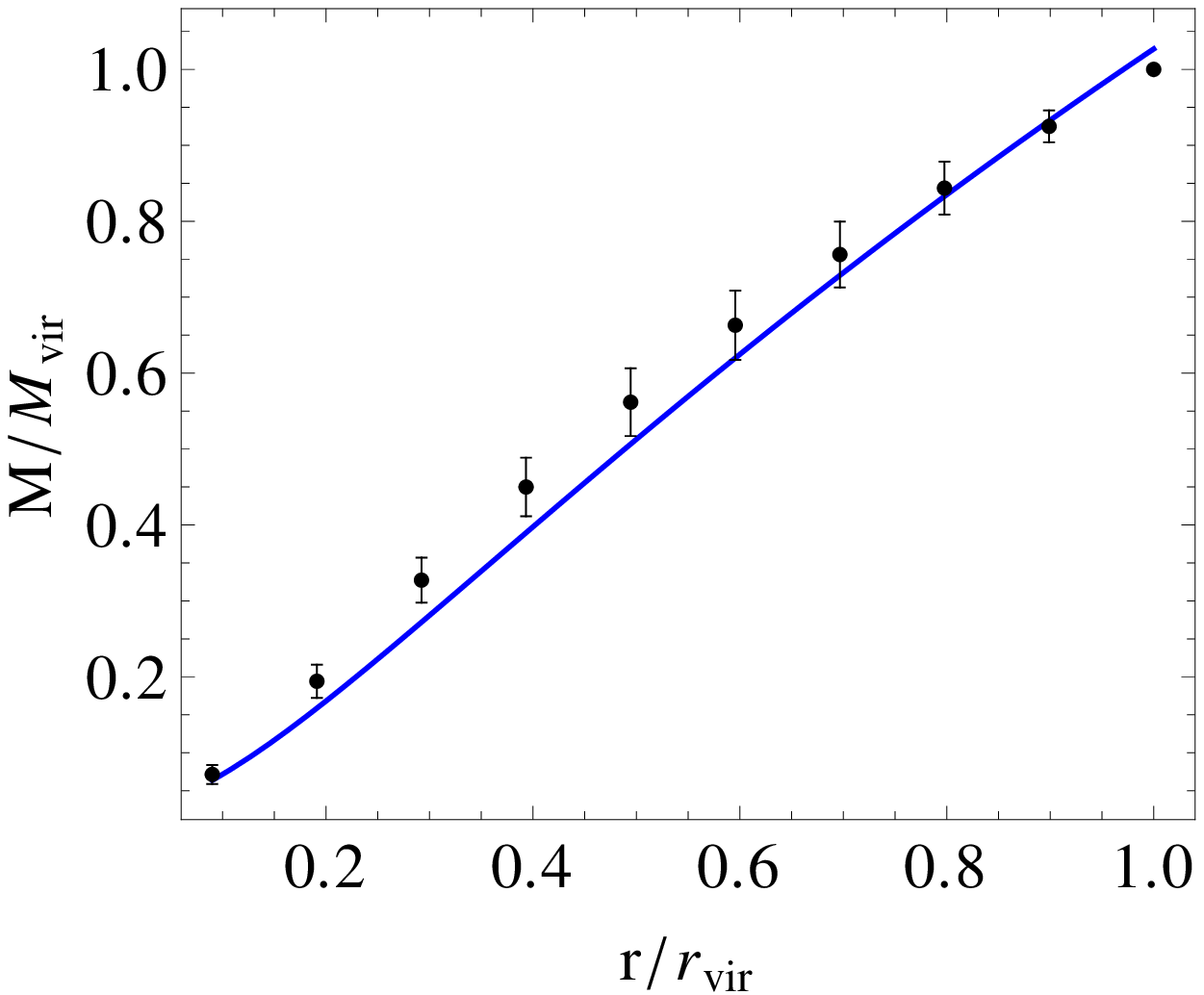}\resizebox{0.73\hsize}{!}{\includegraphics{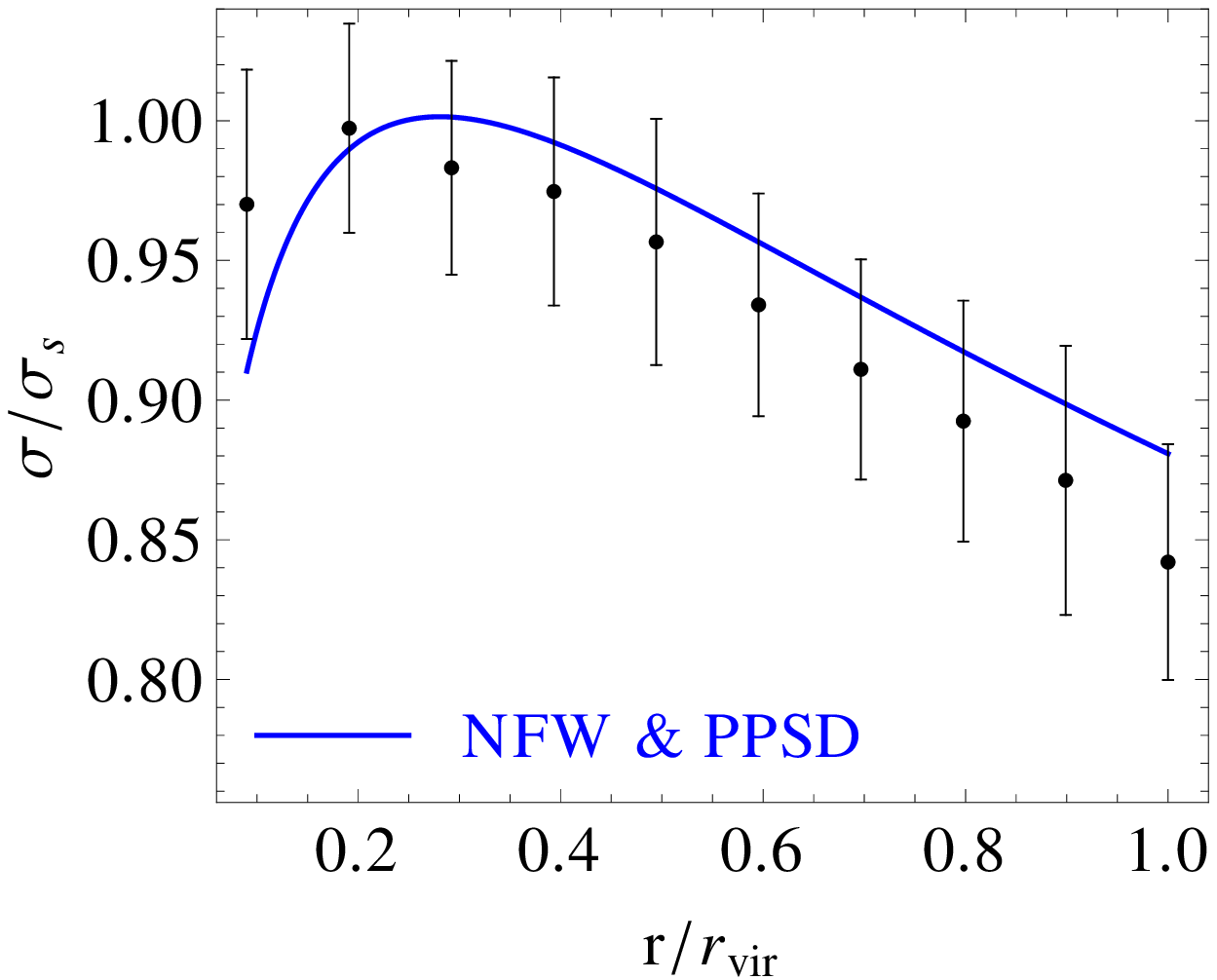}}}
\resizebox{\hsize}{!}{\includegraphics{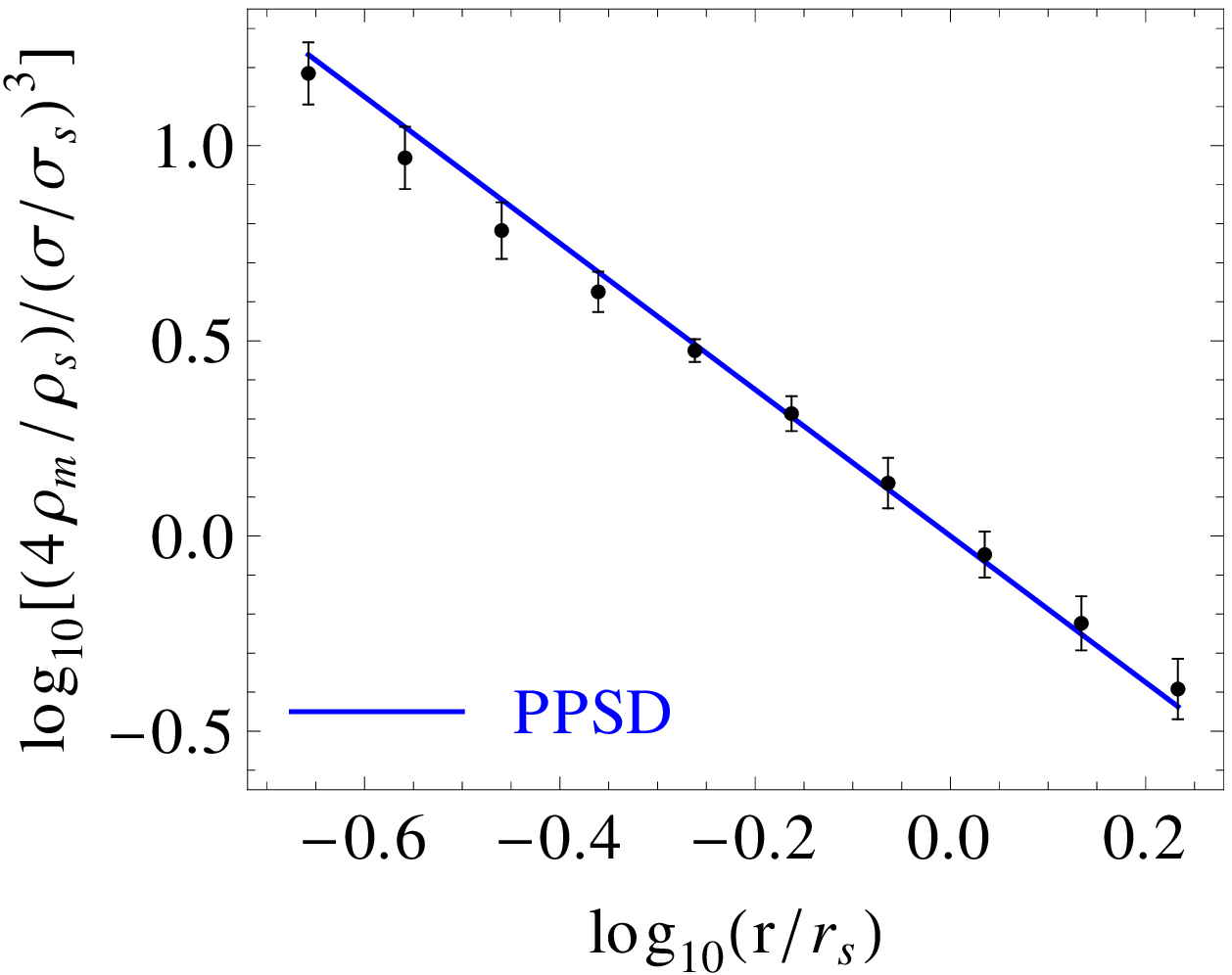}\resizebox{0.69\hsize}{!}{\includegraphics{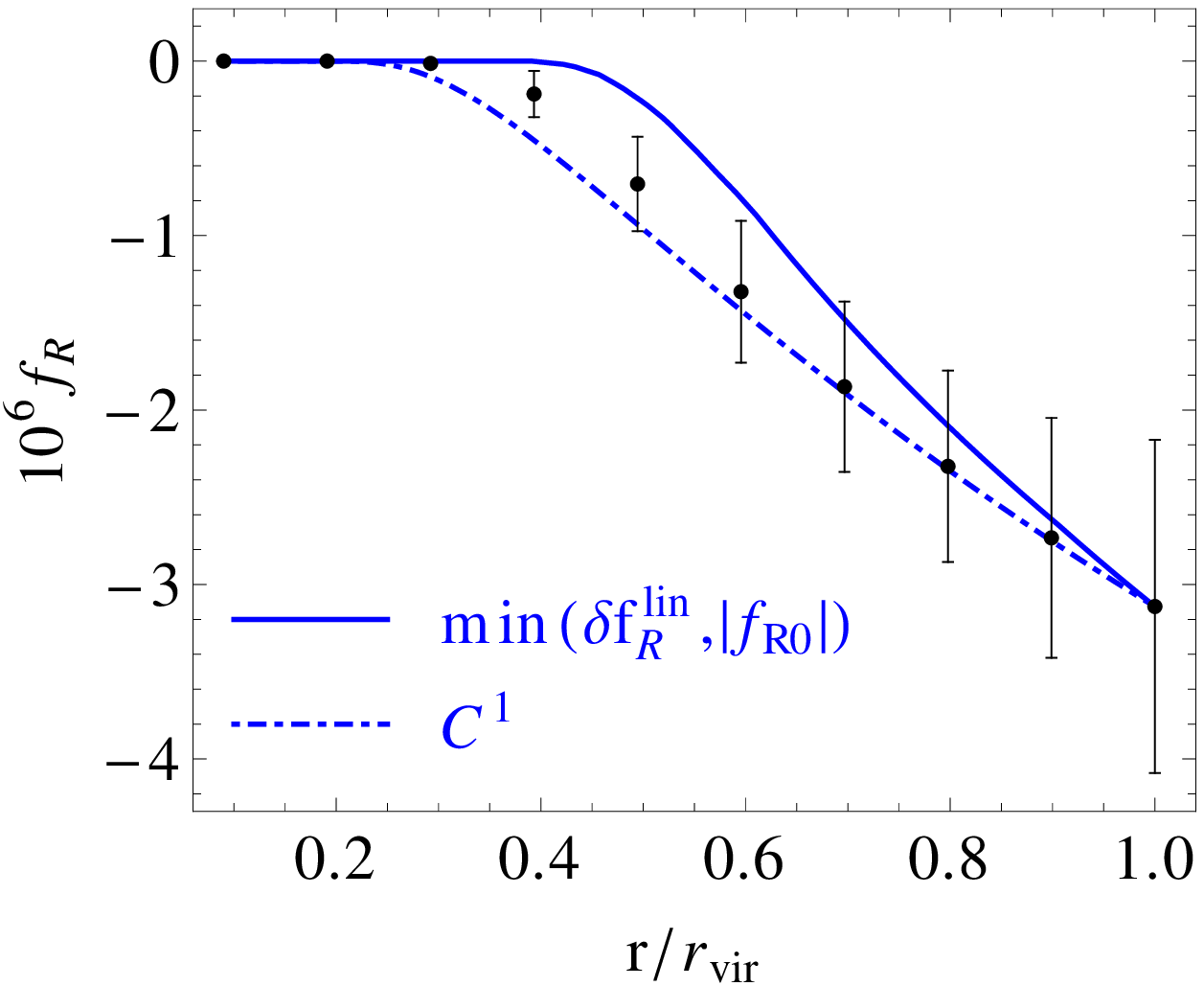}}\includegraphics{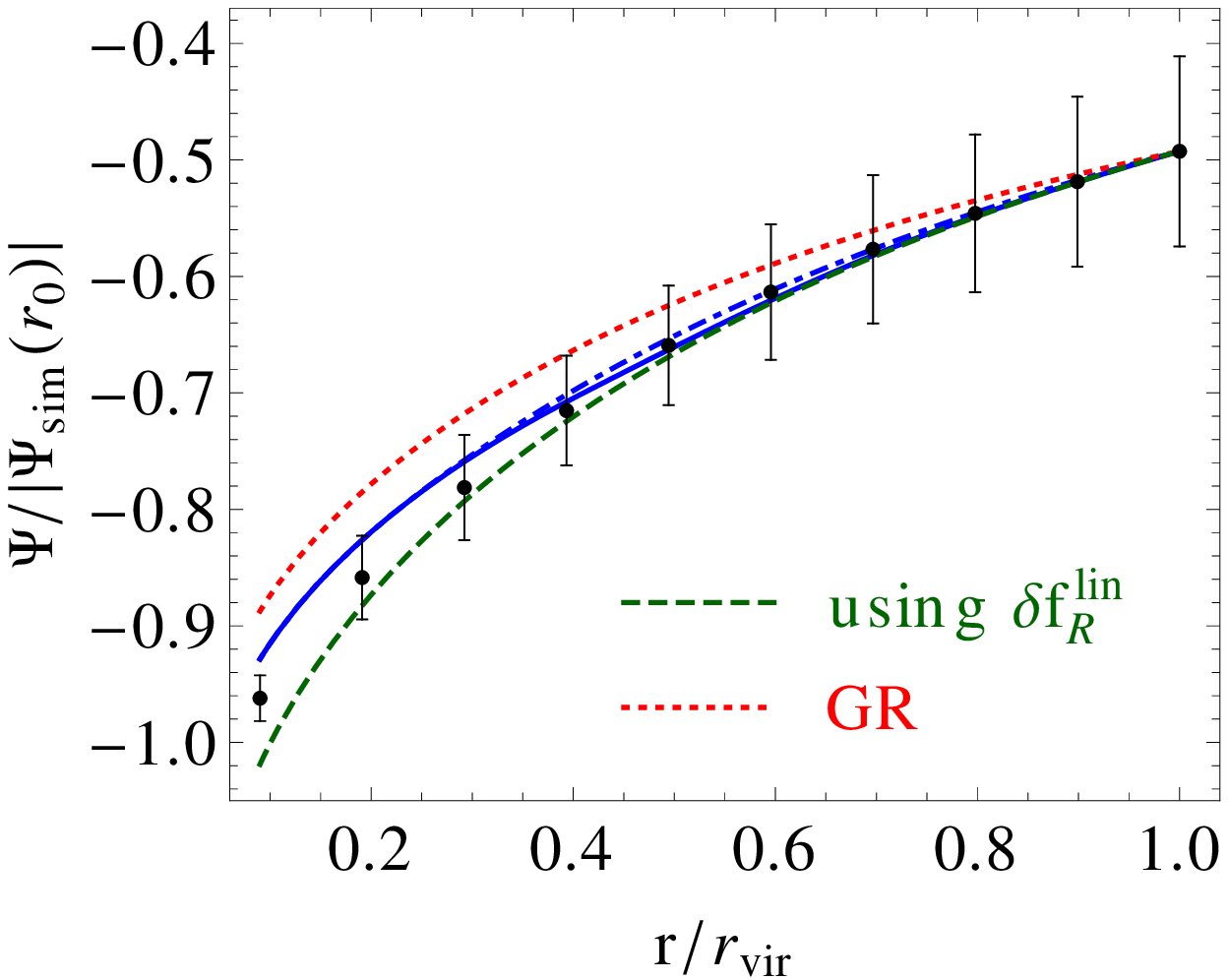}}
\caption{Comparison of radial dependencies of $f(R)$ halo quantities predicted by simulations and the fits constructed in~\textsection\ref{sec:clusterproperties} for $\absfR=10^{-5}$ $(n=1)$ and $L_{\rm Box}=128~\Mpch$ (F-128-5). The simulated halos and their individual fits are stacked for $M=(1.65-1.70)\times10^{14}~\Msunh$. The error bars show the standard deviation in the simulation output. \emph{Top left}: cluster density profile $\drhom/\brhom$. \emph{Top middle}: halo mass $M$. \emph{Top right}: velocity dispersion $\sigma$. \emph{Bottom left}: pseudo-phase-space distribution $\rho/\sigma^3$ normalized at $\rs$. \emph{Bottom middle}: $f_R$ scalar field for the stacked instantaneous (solid) and $\mathcal{C}^1$ (dot-dashed) transition to the chameleon solution. \emph{Bottom right}: gravitational potential $\Psi$ for the $\dfR$ solutions in the bottom middle panel along with the limiting assumptions of Newtonian/GR (dotted) and linearized $f(R)$ (dashed) forces.}
\label{fig:shape}
\end{figure*}

\begin{figure*}
 \resizebox{\hsize}{!}{\includegraphics{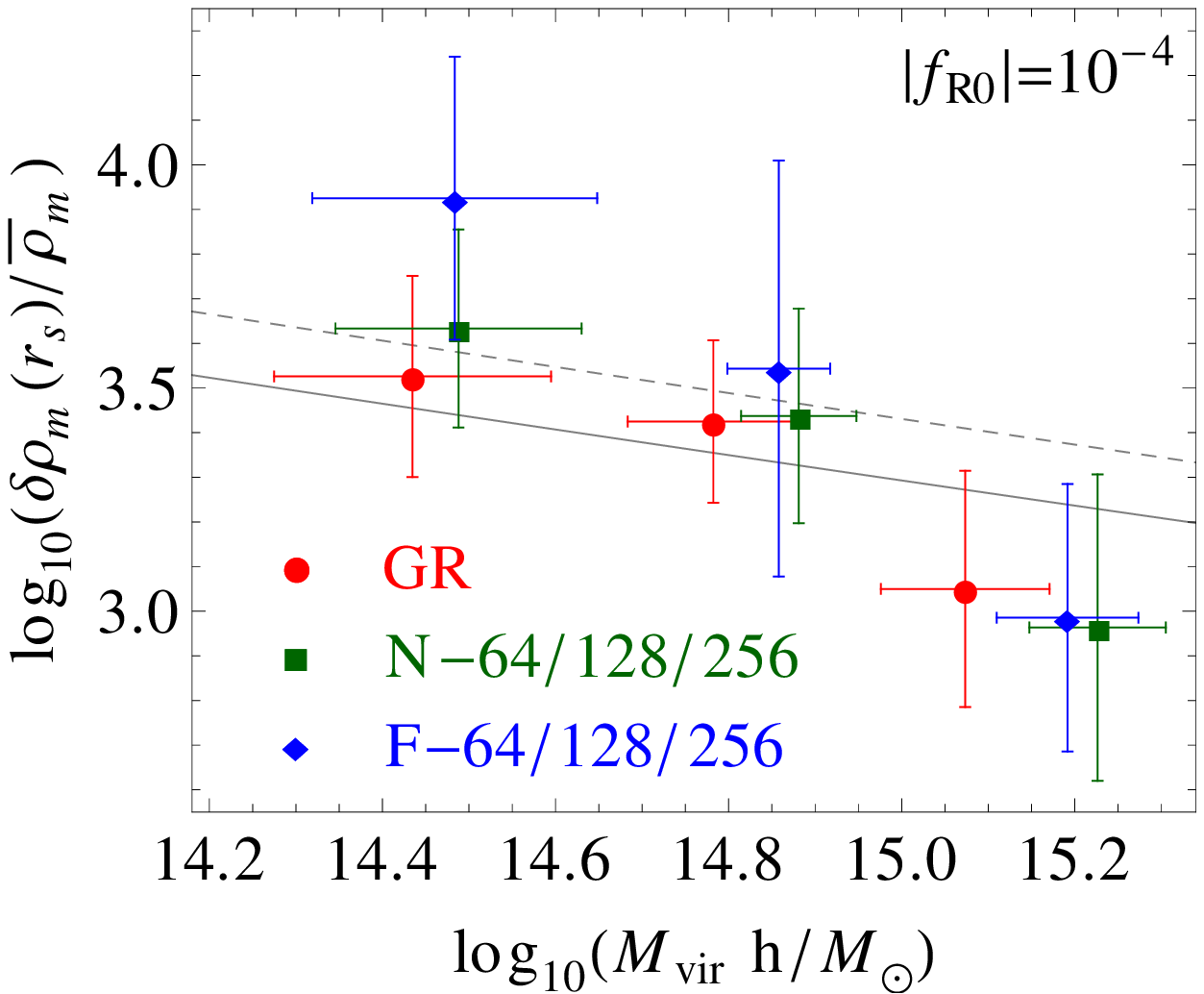}\includegraphics{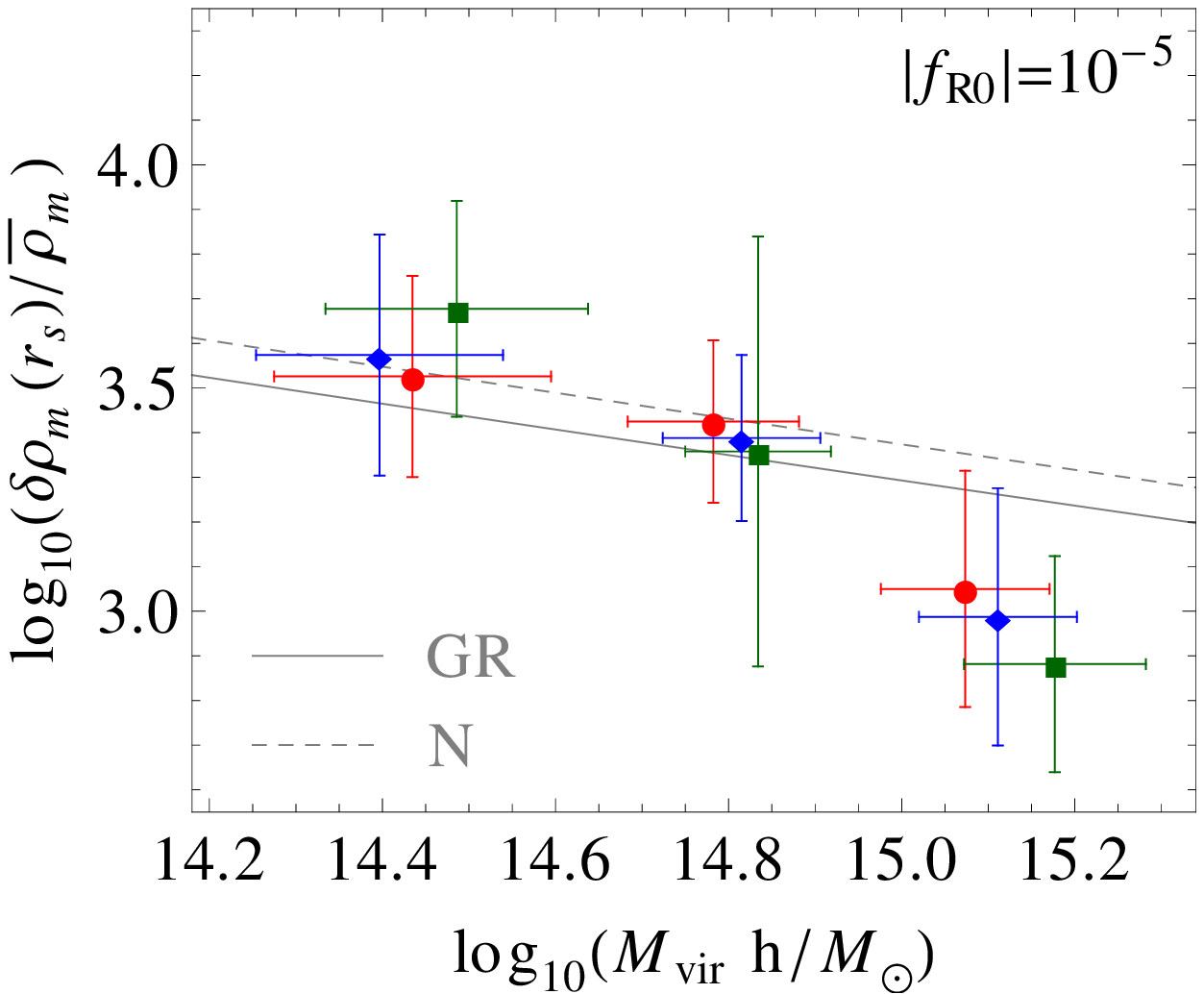}\includegraphics{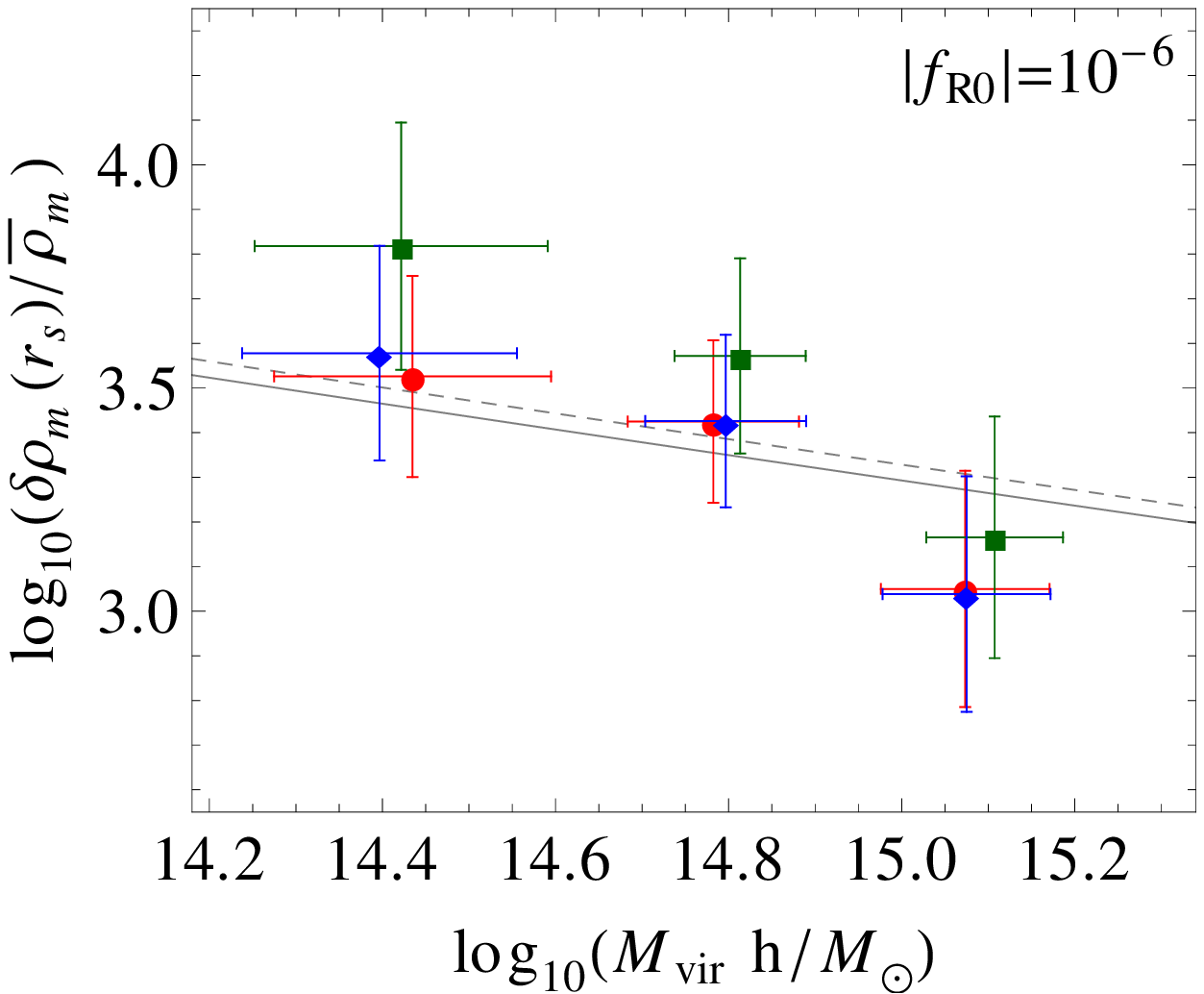}}
 \resizebox{\hsize}{!}{\includegraphics{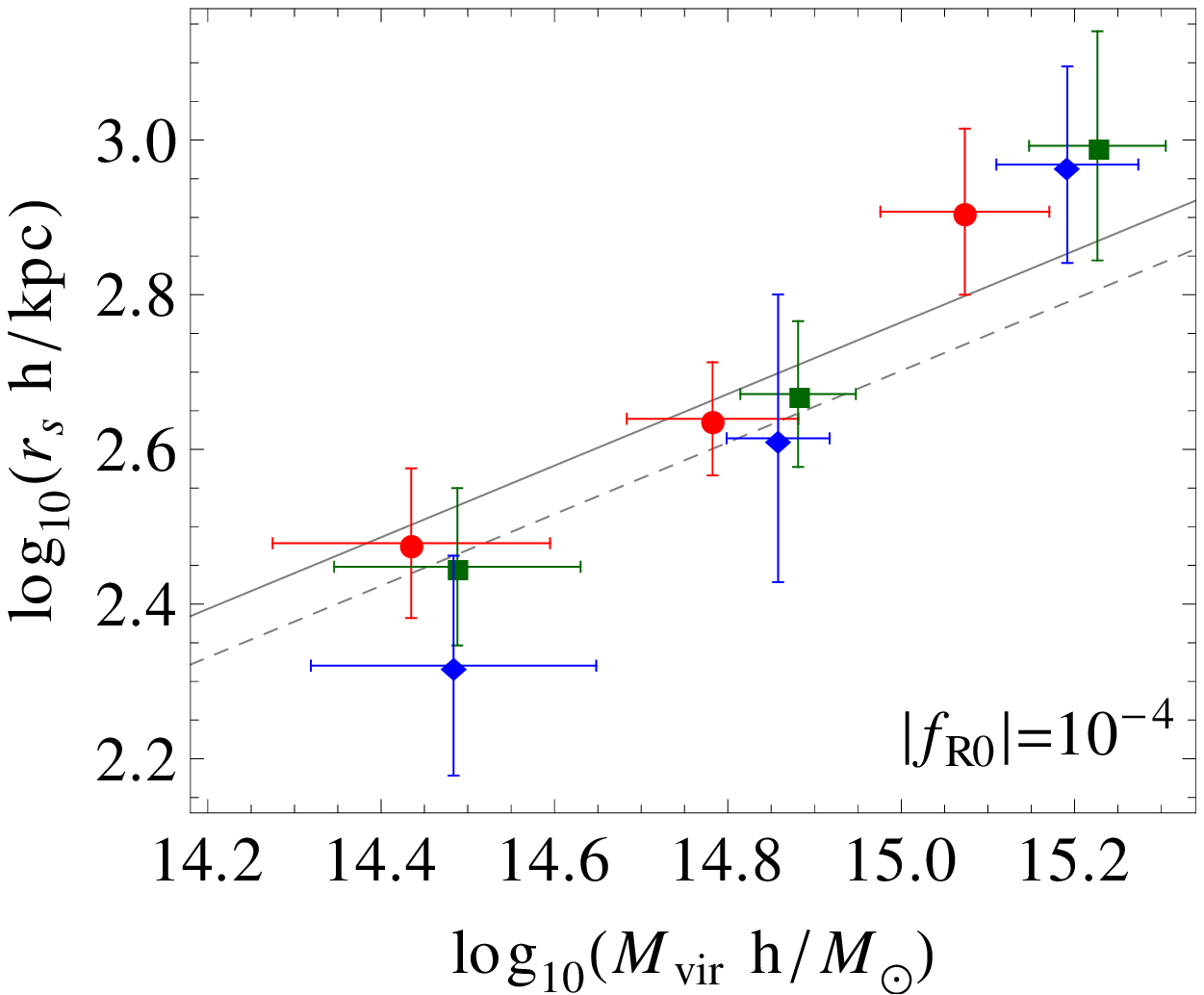}\includegraphics{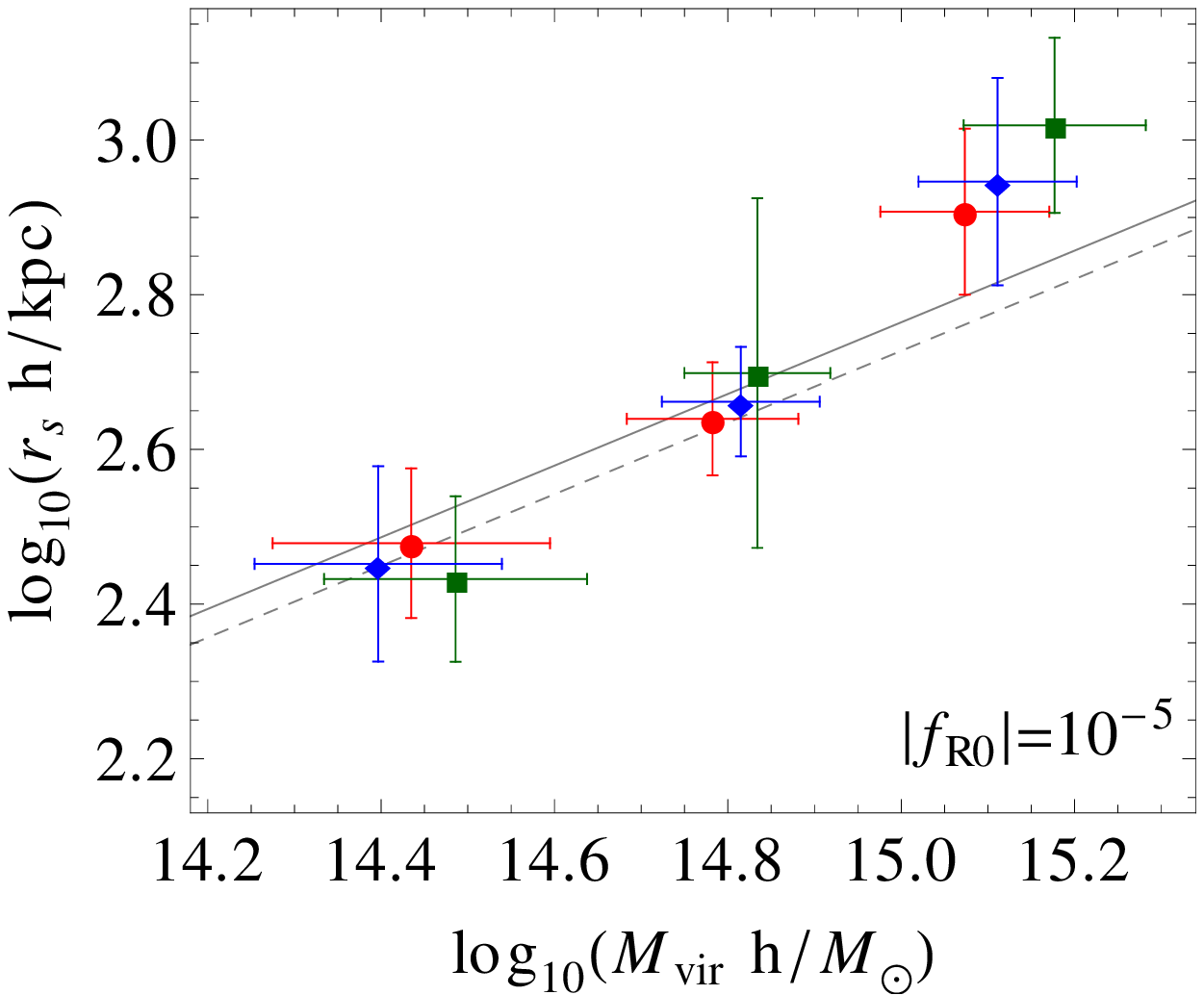}\includegraphics{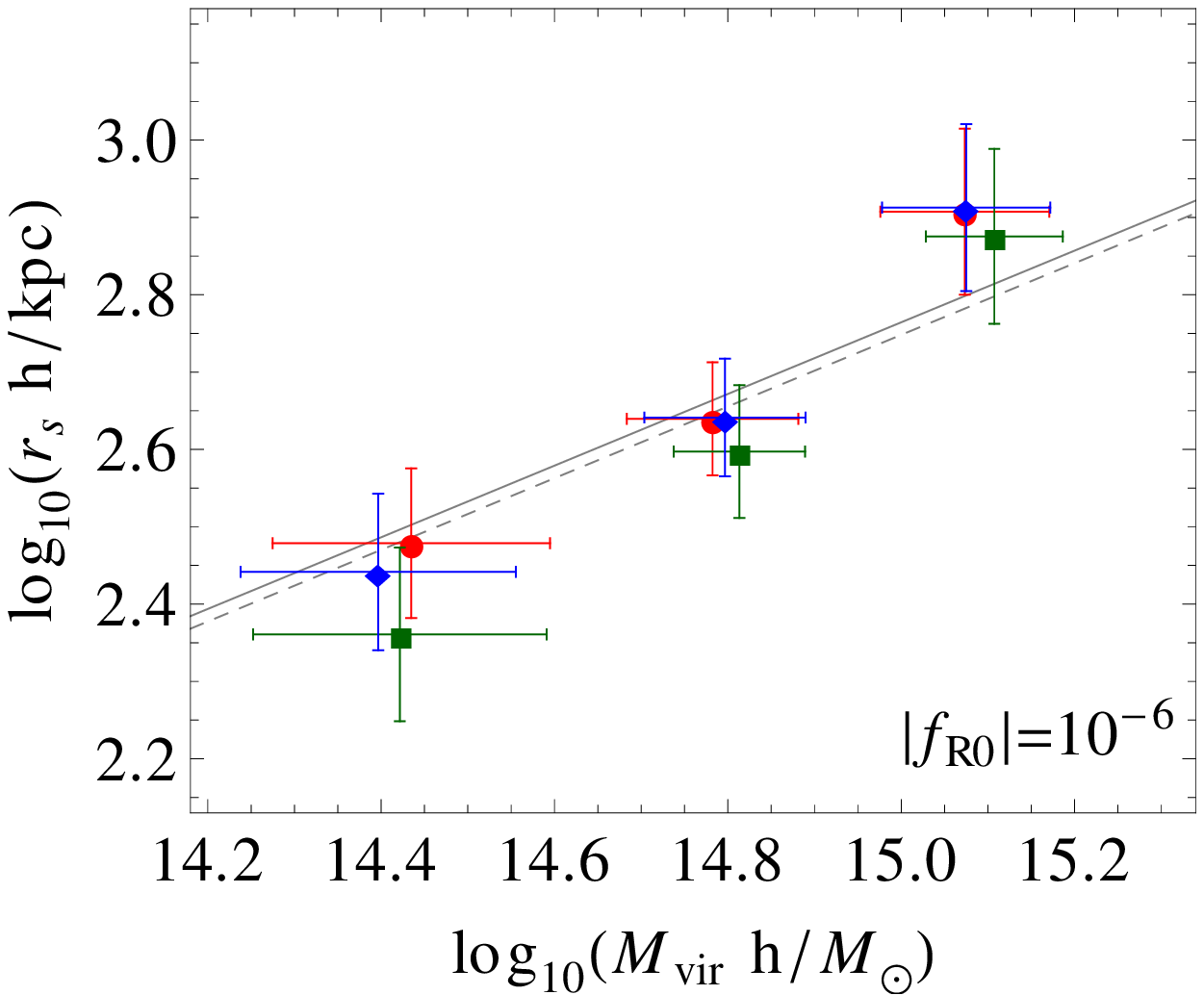}}
 \resizebox{\hsize}{!}{\includegraphics{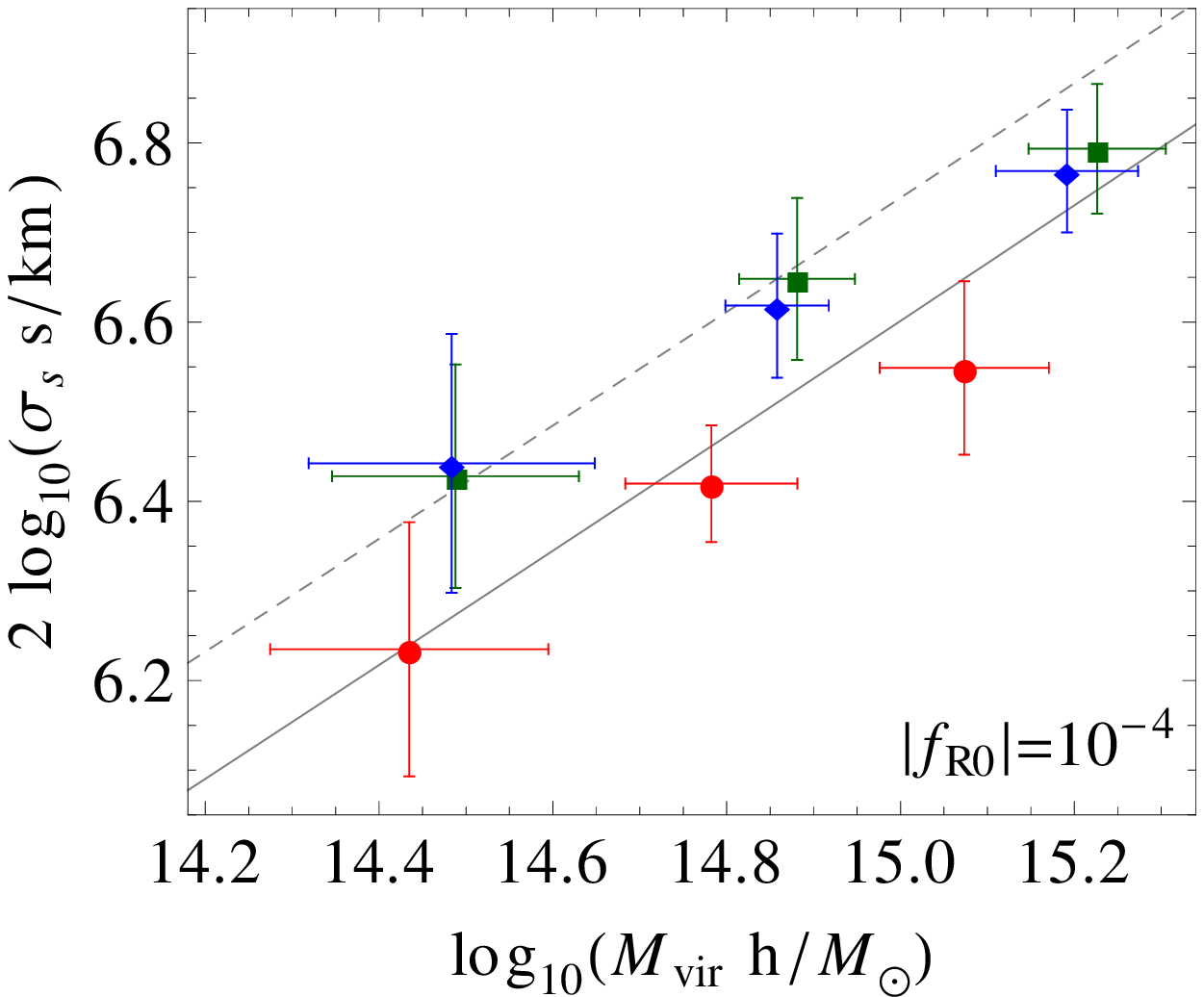}\includegraphics{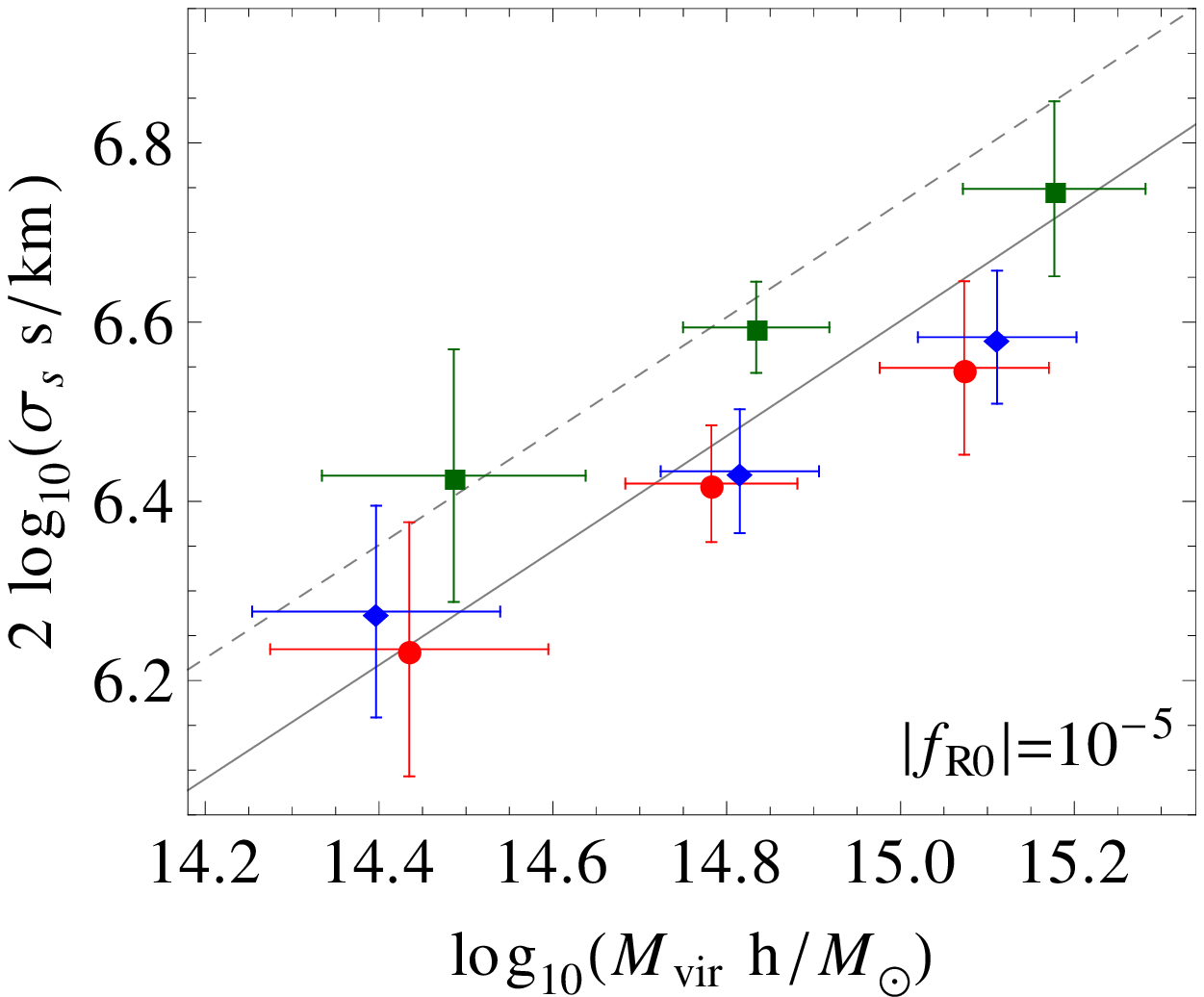}\includegraphics{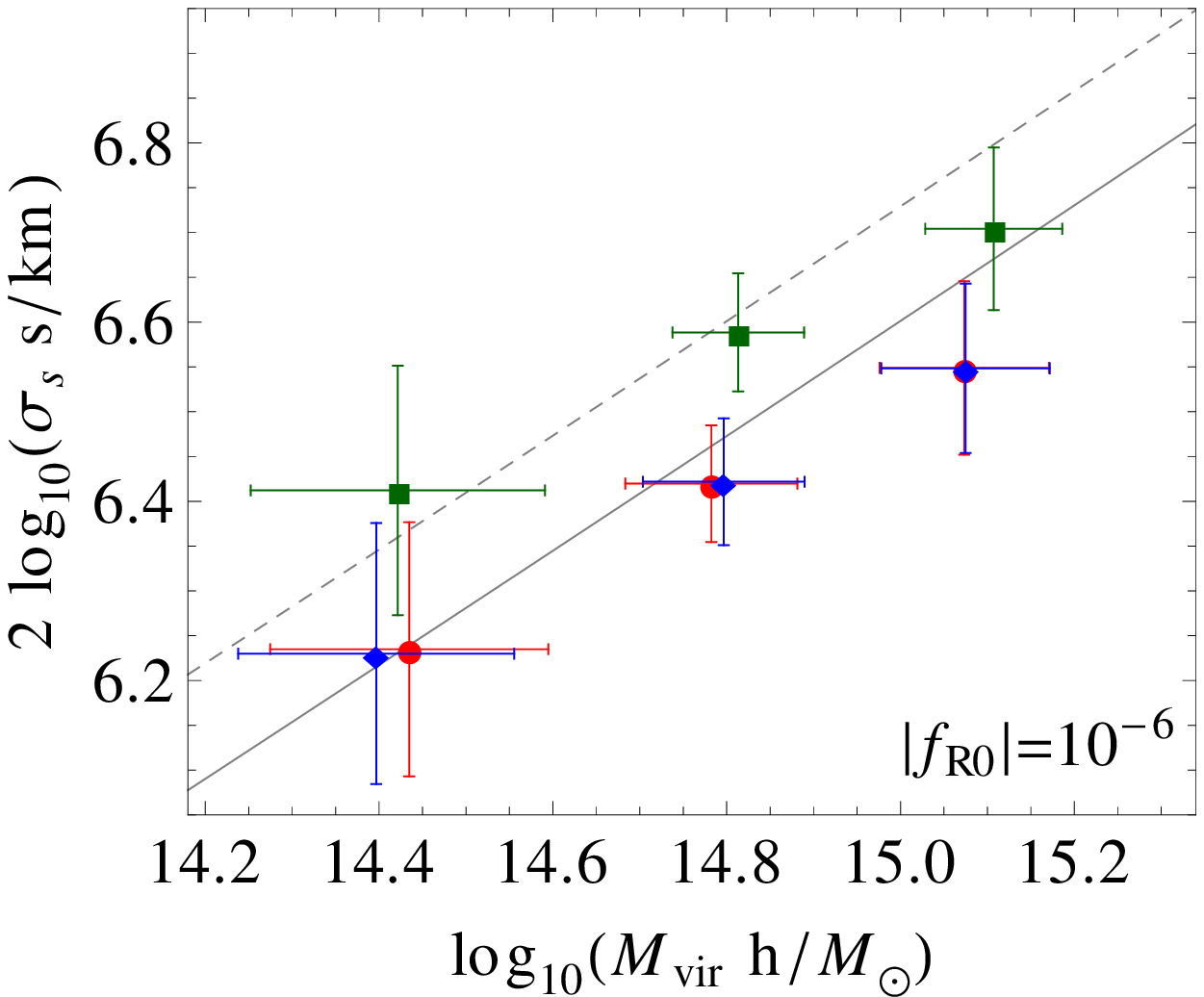}}
 \resizebox{\hsize}{!}{\includegraphics{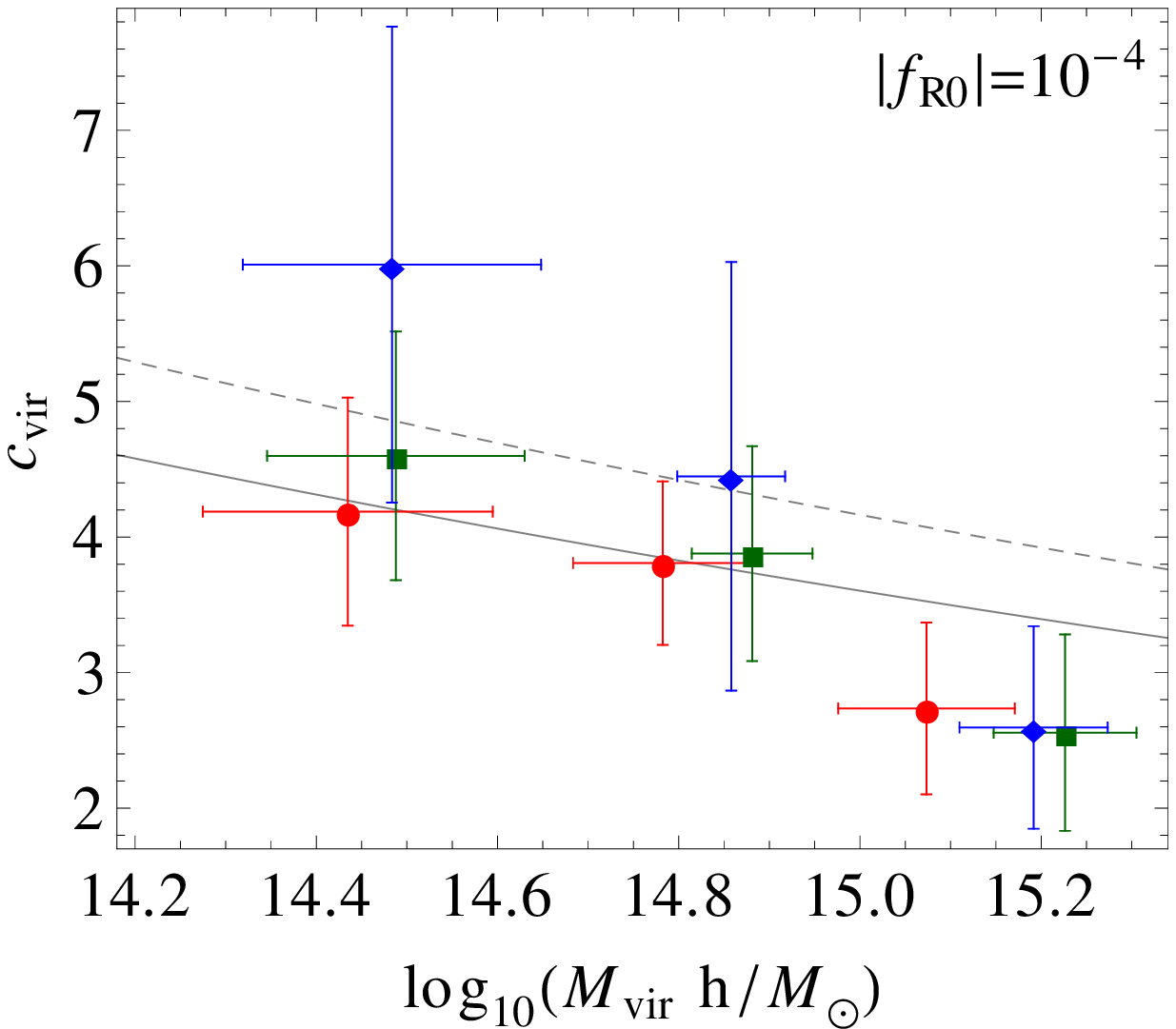}\includegraphics{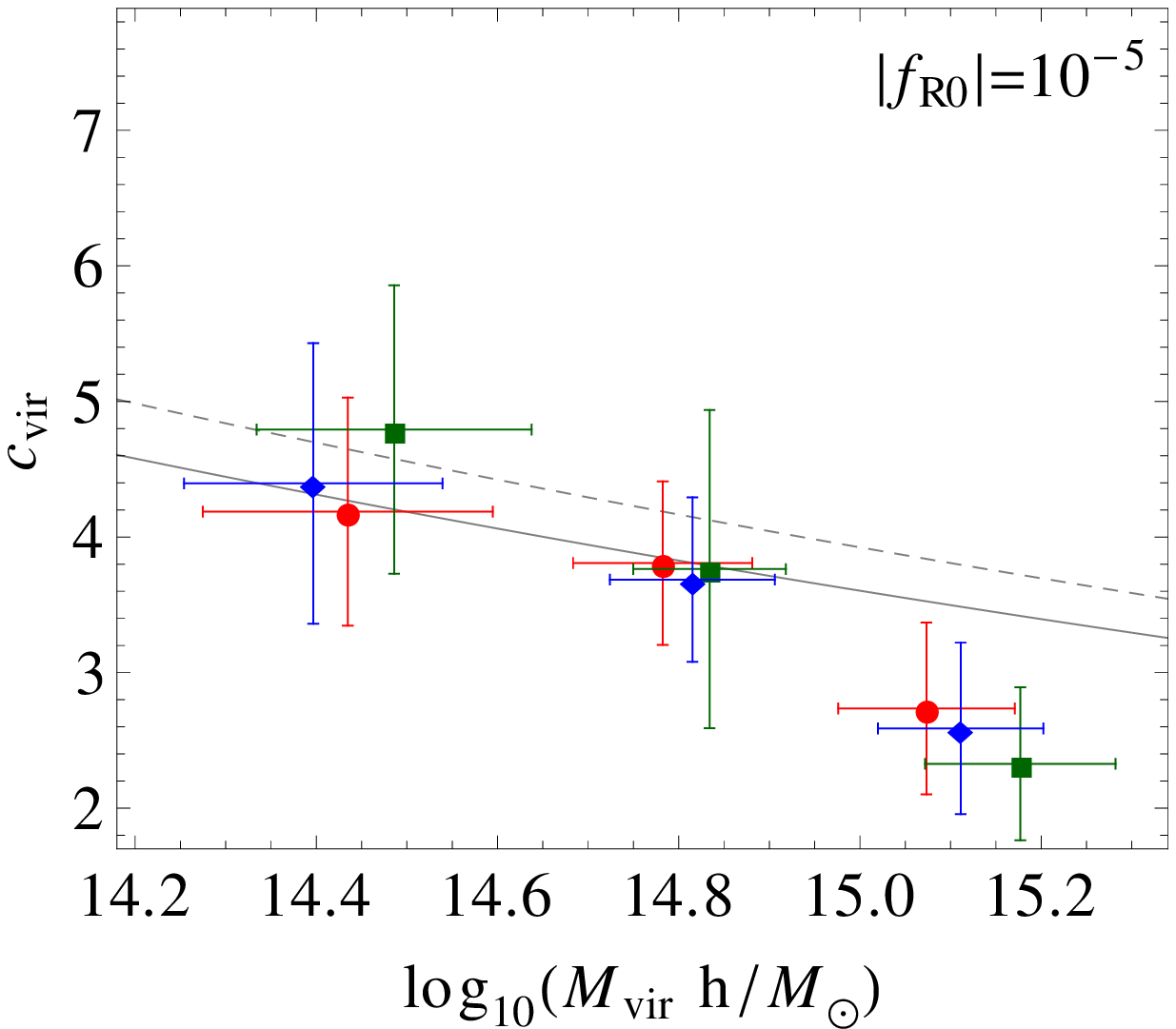}\includegraphics{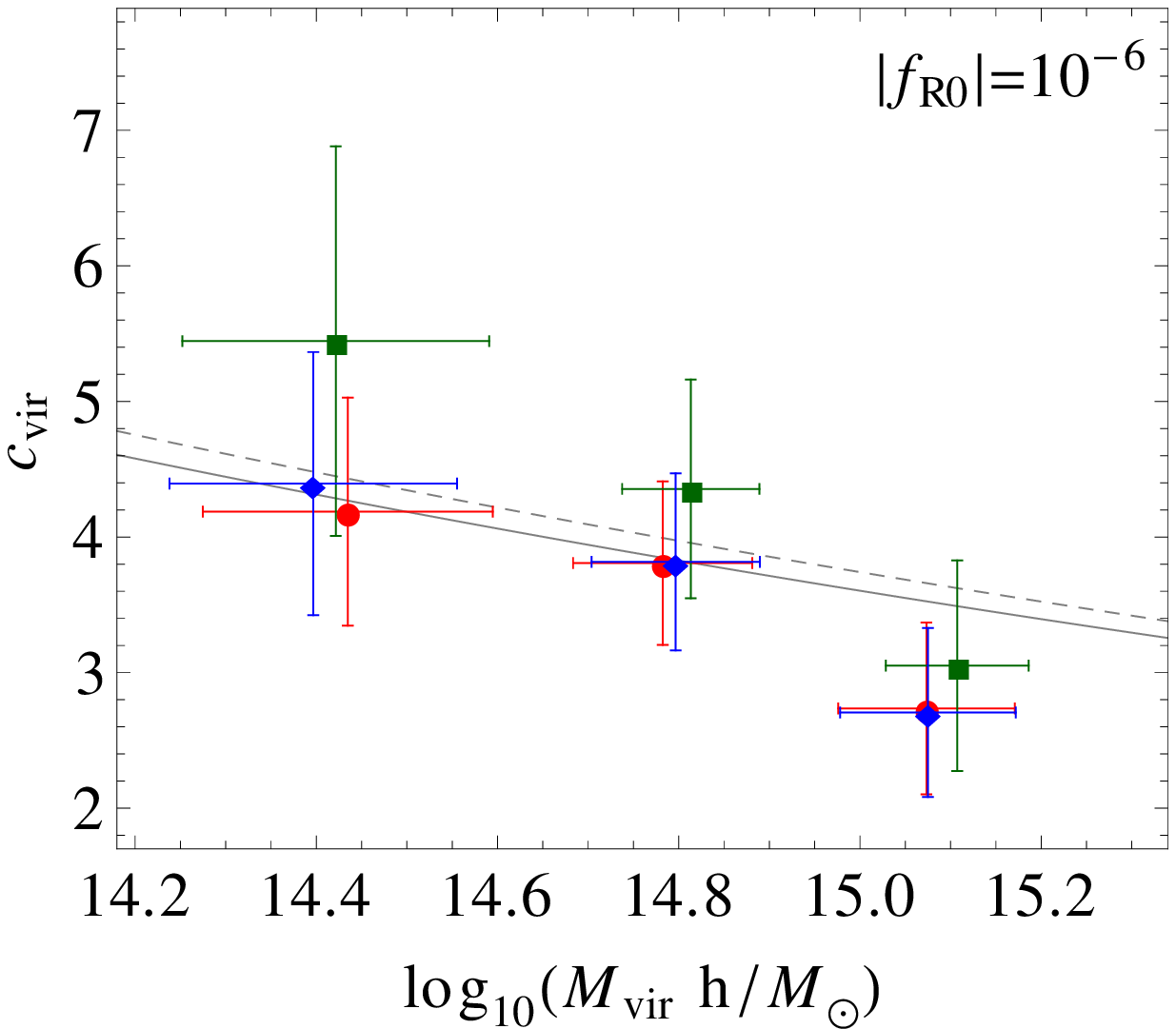}}
\caption{Stacked halo fitting parameters along with the concentration $\cvir$ for the three different box sizes and predictions for them from scaling relations (see Appendix~\ref{sec:scaling}). In $f(R)$ gravity, halo abundances and therefore mean masses are enhanced. Clearly identifiable, in the linearized $f(R)$ regime, velocity dispersions squared are enhanced by a factor of 4/3 over the Newtonian results, i.e., in addition to smaller effects of different $\cvir$. The chameleon effect returns the cluster abundance at high halo masses and the velocity dispersion to the Newtonian/GR predictions.
The predictions from scaling relations based on the spherical collapse and the Jeans function are shown for the Newtonian (GR - solid line) and linearized $f(R)$ (N - dashed line) case, respectively.
\emph{Rows from top}: The matter overdensity at $\rs$ (given by $\beta$ or $\rhos$), the characteristic scale $\rs$, and the velocity dispersion squared at $\rs$, $\sigmas^2$, as well as the concentration $\cvir$. \emph{Columns from left}: $\absfR=10^{-4},\ 10^{-5},\ 10^{-6}$ $(n=1)$.}
\label{fig:coll}
\end{figure*}

We test the predictions made in~\textsection\ref{sec:clusterproperties} on the $z=0$ simulation output described in~\textsection\ref{sec:simulations}.
We calculate the reduced $\chi^2$ of the relative deviation between the prediction for the quantity $q$ from our analytic fits and the simulation output, i.e.,
\bq
 \chi^2_{q,{\rm red}} = \frac{1}{\nu} \sum_{n=1}^N \left( \frac{q_{\rm fit}}{q_{\rm sim}} - 1 \right)^2 \label{eq:chi2red}
\eq
for each halo of the simulation independently, where $N$ is the number of bins in $r\in(r_0,\rvir)$ and $\nu=N-n-1$, $n$ being the number of fitting parameters used in the fit.
We then calculate the mean of Eq.~(\ref{eq:chi2red}) over all halos.
Our results are summarized in Table~\ref{tab:deviations}.
The mass range chosen for the selection of the clusters (see Table~\ref{tab:deviations}) picks about 40-50 halos out of the 50-60 most massive halos of each simulation.
The average is then taken over the 10 realizations of each simulation configuration, leading to an average over about 500 halos for the results shown in each row of Table~\ref{tab:deviations}.
We chose this simple approach of quantifying the goodness of fit only for the qualitative comparison between its performance in the concordance model and in the linearized and chameleon $f(R)$ gravity models.

In particular, we find that both the NFW profile and the power-law PPSD profile with standard radial dependence, $r^{-15/8}$, yield equally good descriptions to the halos produced in the $f(R)$ models as they do in the concordance model.
Note that for the scalar field $\dfR$, we only take into account the instantaneous transition to the chameleon region (see~\textsection{\ref{sec:scalarfield}}) with the mass calibration at $r_0$, Eq.~(\ref{eq:mass}), in correspondence to the mass correction in the gravitational potential Eq.~(\ref{eq:GRpot}).

For illustration of our fits on the simulation data, we choose a set of simulations that highlights the effect of the chameleon mechanism, i.e.,
where the transition from the large-field to the small-field limit takes place within the scales of interest.
Thus, we seek a combination of medium field strength $\absfR$ and medium halo masses.
We therefore illustrate the halo quantities produced for $\absfR=10^{-5}$ and $L_{\rm Box}=128~\Mpch$, corresponding to the set of simulation outputs denoted by F-128-5 (see~\textsection\ref{sec:simulations}).
Fig.~\ref{fig:shape} shows the stacked fits to the F-128-5 simulation output, which is also stacked, for the halo density profile, the cluster mass, the velocity dispersion, the PPSD profile, the scalar field $f_R$, and the gravitational potential. The normalizations and narrow mass range, $(1.65-1.70)\times10^{14}~\Msunh$, are chosen such that standard deviations of the simulation output are small, in particular, at scales where the chameleon mechanism is active.
The narrow mass range is applied to the 10 realizations of the F-128-5 configuration, resulting in taking the average over 16 halos.

In Fig.~\ref{fig:coll} we show the overdensity $\drhom/\brhom$ at $\rs$, the characteristic scale $\rs$, the velocity dispersion squared at $\rs$, $\sigmas^2$, and the concentration parameter $\cvir=\rvir/\rs$ as a function of the virial mass $\Mvir$.
The three types of data points correspond to the Newtonian, linearized $f(R)$, and full chameleon $f(R)$ model, respectively.
The three bins of each type indicate the stacked best-fit values to each halo produced in the simulations of the three different box sizes, where the least massive ones correspond to $L_{\rm Box}=64~\Mpch$ and the most massive ones to $L_{\rm Box}=256~\Mpch$.
We stack the about 10 most massive halos in the selection for Table~\ref{tab:deviations} of each realization of each simulation configuration, resulting in the average over about 100 halos for each data point.
Fig.~\ref{fig:coll} is a good demonstration of the chameleon mechanism in several realizations.
For the $f(R)$ models, we clearly observe a shift of the average mass of each data point towards higher masses with respect to GR.
This corresponds to the enhanced abundance of massive halos in $f(R)$ gravity.
Whereas for high values of $\absfR$ the full chameleon simulations approach the linearized simulations, they reproduce the Newtonian simulations at low values of $\absfR$.
We further observe that in the full chameleon simulations, the displacement in the mean mass with respect to GR is strongest at high masses for large $\absfR$ and at low masses for small $\absfR$, respectively.
This coincides with the expectations of the $f(R)$ halo abundance, i.e., where chameleon simulations recover GR at high masses for low values of $\absfR$ but differ at low masses; and for large values of $\absfR$, the strongest difference to GR is observed at high halo masses (see~\cite{schmidt:08, zhao:10b}).
A further realization of the chameleon effect can be observed in the square of the velocity dispersion.
Here, the predictions of the chameleon simulations correspond to the linearized simulations for $\absfR=10^{-4}$ that are enhanced by a factor of $\sim 4/3$ over the GR predictions. For $\absfR \lesssim 10^{-5}$, however, the chameleon simulations recover the GR velocity dispersion as the $f(R)$ modification is suppressed and the gravitational force returns to being Newtonian.

\subsection{Prediction of fitting parameters}

We predict the virial halo concentration $\cvir$ along with the two fitting parameters $\rhos$ and $\rs$ in $f(R)$ gravity and GR using scaling relations based on spherical collapse calculations (see Appendix~\ref{sec:scaling}) following~\cite{schmidt:08}.
In order to determine the third fitting parameter, the velocity dispersion at the characteristic scale $\rs$, $\sigmas$, we require that the Jeans equation must be satisfied at $\rs$ given the assumptions of a NFW halo density profile and the standard radial dependence of the PPSD profile, along with a fit for the velocity dispersion anisotropy relation based on concordance model simulations (see Appendix~\ref{sec:scaling}).

We find that the scaling relations obtained in this way yield qualitatively good reproductions of the best-fit values of $\rhos$, $\rs$, $\sigmas$, and $\cvir$.
Our results are shown in Fig.~\ref{fig:coll}.

\section{Conclusions} \label{sec:conclusions}

Modifications of gravity have extensively been tested on solar-system scales (see, e.g.,~\cite{will:05}) and to a lesser degree at large cosmological scales using specific alternative theories of gravity (e.g.,~\cite{fang:08a, lombriser:09, reyes:10, lombriser:10, schmidt:09}), as well as generic modifications to GR while adopting a $\Lambda$CDM background (e.g.,~\cite{rapetti:09, bean:10, daniel:10, zhao:10, dossett:11b,brax:11}) or simultaneously allowing a dynamic effective dark energy equation of state~\cite{lombriser:11, zhao:11}.
However, gravity may also be tested by the structure observed at intermediate scales~\cite{smith:09t, wojtak:11, lombriser:11b}.
In this regime, nonlinear gravitational interactions gain in importance and need to be modeled correctly to obtain reliable predictions for both GR and its competitors, which in turn can be compared with observations to infer constraints on modified gravity theories.
Besides the usual difficulties of modeling the nonlinear structure known to studies of the standard Newtonian gravity, modifications of gravity may be complicated by additional nonlinear mechanisms such as the chameleon effect, which suppresses modifications of gravity in high density regions.
In order to obtain reliable constraints from observations, these effects need to be consistently incorporated.

In this paper, we concentrate on $f(R)$ gravity and aim at describing the scalar field, the gravitational potential, and the velocity dispersion within virialized clusters produced in $f(R)$ $N$-body dark matter simulations of the Hu-Sawicki model.
We derive and test analytic, semi-analytic, and numerical relations for these quantities that can be used to compare theory with observations at the virialized scales of clusters.
We assume the standard NFW halo density profile and the PPSD profile with usual power law, which we find to provide comparably good fits to the $f(R)$ scenario as they do for the concordance model.
We argue that this is not unexpected from the consideration of modified forces in the secondary infall of a collisional gas for the approximate description of the $N$-body collisionless dark matter system.
The fits to the simulation output are based on three degrees of freedom, the characteristic density $\rhos$, the characteristic scale $\rs$, and the velocity dispersion at $\rs$, $\sigmas$.
We find that scaling relations based on the gravitational collapse and the requirement of the validity of the Jeans equation yield good qualitative predictions for these quantities when accounting for the modified forces at work.

The extension of our results to scales beyond the virial radius exhibits additional challenges, such as the correct modeling of the two-halo contribution (cf.~\cite{schmidt:08,lombriser:11b}), which we shall leave for future work.

\section*{Acknowledgments}

We thank Francesco Pace and Marc Manera for useful discussions.
LL and KK are supported by the European Research Council and KK and GBZ are supported by the STFC (grant no. ST/H002774/1). KK is also supported by the Leverhulme trust. Numerical computations are done on the Sciama High Performance Compute cluster which is supported by the ICG, SEPnet, and the University of Portsmouth. 

\appendix

\section{Density profiles in modified gravity} \label{sec:selfsiminfall}

In~\textsection\ref{sec:results}, we have seen that the NFW halo density profile for $r\in(r_0,\rvir)$ provides as good fits to the virialized clusters in $f(R)$ gravity as it does in the Newtonian scenario.
In the following, we shall give qualitative arguments for why this may be expected from a theoretical point of view.
In order to study the gravitational collapse in $f(R)$ gravity,
we parametrize the enhancement of gravitational forces by the factor $(1-F)$, where for simplicity, the quantity $F$ is defined by
\bq
 F = \left\{ \begin{array}{lll} -1/3, & f(R) \ \textrm{gravity}, \\ 0, & \textrm{Newton/GR}, \end{array} \right.
\eq
which holds in the linear regime and when modifications are absent or suppressed, respectively (see right panel of Fig.~\ref{fig:scalaron}).

\subsection{Self-similar infall of a collisional gas} \label{sec:selfsiminfallcollgas}

We follow Bertschinger~\cite{bertschinger:85} for the self-similar infall and the shocked accretion of a collisional gas onto the center of an initially spherical uniform overdensity in an otherwise uniformly expanding Einstein-de Sitter universe.
The equations governing the postshock motion of the fluid are, nondimensionalized (see~\cite{bertschinger:85}),
\bqa
 (V - \frac{8}{9} \lambda) D' + D \, V' + \frac{2D \, V}{\lambda} - 2D & = & 0, \nonumber \\
 (V - \frac{8}{9} \lambda) V' - \frac{1}{9} V + \frac{P'}{D} & = & - \frac{2}{9} \frac{M}{\lambda^2} (1-F), \nonumber \\
 (V - \frac{8}{9} \lambda) \left( \frac{P'}{P} - \gamma \frac{D'}{D} \right) & = & \frac{20}{9} - 2 \gamma, \nonumber \\
 M' & = & 3 \lambda^2 D,
 \label{eq:nondimensionalized}
\eqa
i.e., the continuity, Euler, adiabatic, and mass equations.
Here, $V$, $D$, $P$, and $M$ is the nondimensionalized velocity, density, pressure, and mass, respectively, and $\gamma$ indicates the ratio of specific heats, which is taken to be $\gamma=5/3$, i.e., the ratio for a monatomic ideal gas.
Primes denote derivatives with respect to $\lambda=r/\rta$, where $\rta$ is the turn-around radius.
Note that we have introduced here the modification of the gravitational force $F$.
The nondimensionalized quantities relate to the fluid variables via (see~\cite{bertschinger:85})
\bqa
 v(r,t) & = & \frac{\rta}{t} V(\lambda), \nonumber \\
 \rho(r,t) & = & \rhoEdS D(\lambda), \nonumber \\
 p(r,t) & = & \rhoEdS \left( \frac{\rta}{t} \right)^2 P(\lambda), \nonumber \\
 m(r,t) & = & \frac{4\pi}{3} \rhoEdS \rta^3 M(\lambda),
\eqa
where $\rhoEdS = 4 \kappa^{-2} t^{-2}/3$ is the Einstein-de Sitter background density and $t$ is the cosmic time.

Factoring out the asymptotic behavior of Eqs.~(\ref{eq:nondimensionalized}) at the origin, requiring the boundary conditions
\bq
 V = M = 0, \ \ \ \ \lambda=0,
\eq
characterizing the shock,
it is easy to see that
\bq
 D = \lambda^{-9/4} \tilde{D}(\lambda), \ \ \ \ P = \lambda^{-5/2} \tilde{P}(\lambda), \ \ \ \ M = \lambda^{3/4} \tilde{M}(\lambda)
 \label{eq:ssinfallres1}
\eq
with finite $\tilde{D}$, $\tilde{P}$, and $\tilde{M}$ at $\lambda=0$.
In the following, we assume that $\lambda$ (or $r$) is sufficiently close to the origin.
Note that changing $F$, changes the relation between $\tilde{P}$ and $\tilde{D}$ too, but
dependencies on $\lambda$ remain the same, i.e., the nondimensionalized radial dependence of the density profile is not affected by the force modification.
Therefore, for equal nondimensionalized mass,
\bq
 \tilde{D}=\tilde{D}_{\rm GR}, \ \ \ \ \tilde{P}=\tilde{P}_{\rm GR}(1-F).
\eq
We compare halos in the different gravitational models, however, by equating the corresponding virial (dimensional) masses with each other.
But since we define the virial radius by the same virial overdensity $\Delta_{\rm vir}$ in all of the models (see Appendix~\ref{sec:scaling}), assuming equivalent background, the virial radius $\rvir$ and therefore $\rta$ are the same.
For a gas of pressure $p=\rho \, \sigma^2$, this therefore implies
\bq
 \rho(r) = \rho_{\rm GR}(r), \ \ \ \ \sigma(r)^2=\sigma_{\rm GR}(r)^2 (1-F).
 \label{eq:ssinfallres2}
\eq

Note, however, that Eq.~(\ref{eq:ssinfallres1}) implies $\rho\sim r^{-9/4}$, which does not provide a good fit to simulated dark matter halos,
but we shall assume for now that the relations in Eq.~(\ref{eq:ssinfallres2}) hold even in cases where $\rho(r)$ is not described by a simple power law.
As pointed out in~\textsection\ref{sec:jeans} and \textsection\ref{sec:virial}, this assumption remains consistent with the Jeans equation and virial theorem, respectively.
Rather than the directly predicted radial dependence in Eq.~(\ref{eq:ssinfallres1}), we are interested in the relation $D^{5/2}/P^{3/2}$, which defines the PPSD $\rho/\sigma^3$.
According to the relations in Eq.~(\ref{eq:ssinfallres1}),
\bq
 \frac{\rho(r)}{\sigma(r)^3} \propto r^{-15/8},
\eq
which in turn is found to yield a good description for the results obtained from CDM simulations~\cite{taylor:01}.
With the force modification $F$, we find that
\bq
 \frac{\rho(r)}{\sigma(r)^3} = \frac{\rho_{\rm GR}(r)}{\sigma_{\rm GR}(r)^3} \frac{1}{(1-F)^{3/2}}.
 \label{eq:fRpsd}
\eq

Finally, note that we consider collisionless dark matter in this paper, whereas for simplicity, we have assumed here a collisional fluid.

\subsection{Jeans equation} \label{sec:jeans}

The Jeans equation is derived from the collisionless Boltzmann equation, which describes the particle phase-space distribution as a function of position, momentum, and time.
Thereby, the Boltzmann equation is multiplied by and integrated over the velocity.
The collisionless Boltzmann equation is the analog to the conservation of energy-momentum and thus applies to all metric theories of gravity and hence the $f(R)$ model considered here.
In spherical coordinates and with the force enhancement $F$, the Jeans equation can be written as
\bqa
\mathcal{D} \sigma_r^2 & = & -(1-F) \frac{\rmd}{\rmd r} \Psi_{\rm GR}, \nonumber \\
 \mathcal{D} & = & \left(\frac{\rmd}{\rmd r} + \frac{\rmd \ln \rho}{\rmd r} + \frac{2\beta_{\sigma}}{r} \right),
 \label{eq:jeans}
\eqa
where $\sigma_r$ denotes the radial component of the velocity dispersion $\sigma$, $\mathcal{D}$ defines a linear operator, and $\beta_{\sigma}(r)$ denotes the anisotropy in the velocity dispersion, i.e.,
\bq
 \beta_{\sigma} = 1 - \frac{\sigma_{\theta}^2 + \sigma_{\varphi}^2}{2\sigma_r^2}.
\eq
Using Eq.~(\ref{eq:fRpsd}), we infer
\bqa
 \mathcal{D} \, \left[ \left(\frac{\rho}{\rho_{\rm GR}}\right)^{2/3} \sigma^2_{r, {\rm GR}} \right] = -\frac{\rmd}{\rmd r} \Psi_{\rm GR} = \mathcal{D}_{\rm GR} \sigma_{r, {\rm GR}}^2,
\eqa
where we have divided the Jeans equation by $(1-F)$ and $\mathcal{D}_{\rm GR}$ is the linear operator with $\rho_{\rm GR}$ and $\beta_{\sigma, {\rm GR}}$.
Hence, for $\rho=\rho_{\rm GR}$ and $\sigma_r^2=\sigma_{r, {\rm GR}}^2(1-F)$, the Jeans equation is satisfied, implying $\beta_{\sigma} = \beta_{\sigma, {\rm GR}}$.
Note, however, that this solution is not the only one satisfying Eq.~(\ref{eq:jeans}).
Describing the alternative solutions is, however, beyond the scope of this paper.

\subsection{Virial theorem} \label{sec:virial}

Multiplying the collisionless Boltzmann equation by velocity and position and integrating over both, we obtain for a system in steady state the virial theorem
\bq
 W_{\rm GR} = - 2 T_{\rm GR}, \label{eq:virialtheorem}
\eq
where $W_{\rm GR}$ and $T_{\rm GR}$ are the potential and kinetic energies in Newtonian gravity, respectively, i.e.,
\bqa
 W_{\rm GR} & \equiv & -\int d^3{\bf x} \, \rho_{\rm GR}({\bf x}) {\bf x} \cdot {\bf \nabla} \Psi_{\rm GR} ({\bf x}), \label{eq:potenerg} \\
 T_{\rm GR} & \equiv & \frac{1}{2} \int d^3{\bf x} \, \rho_{\rm GR}({\bf x}) \sigma_{\rm GR}^2({\bf x}).  \label{eq:kinenerg}
\eqa
With $\rho=\rho_{\rm GR}$ and $\sigma^2=(1-F)\sigma_{\rm GR}^2$, we have
$W = (1-F) W_{\rm GR}$ and $T = (1-F) T_{\rm GR}$, which consistently reproduces
the virial theorem $W = -2 T$.
Following from the Boltzmann equation, this is as expected since using the Jeans equation in the integration of Eq.~(\ref{eq:potenerg}) or Eq.~(\ref{eq:kinenerg}) leads to the virial theorem, Eq.~(\ref{eq:virialtheorem}).

\section{Scaling relations from spherical collapse and the Jeans equation} \label{sec:scaling}

The three degrees of freedom used in the fits of~\textsection\ref{sec:clusterproperties}, the amplitude $\rhos$ (or $\beta$) of the NFW halo density profile, Eq.~(\ref{eq:nfw}), the characteristic scale $\rs$, and the velocity dispersion at $\rs$, $\sigmas$, can be predicted by scaling relations based on the spherical collapse (see, e.g.,~\cite{schmidt:08}) and the Jeans equation.
Here, we assume the usual collapse density $\delta_{\rm c} \approx 1.673$ and virial overdensity $\Delta_{\rm vir} \approx 390$ inferred from spherical collapse calculations with a standard force, i.e., $F=0$.
See~\cite{schmidt:08} for derivations of $\delta_{\rm c}$ and $\Delta_{\rm vir}$ in the modified spherical collapse with enhanced force $F=-1/3$.

The variance is defined by
\bq
 \sigma_{\rm var}^2(r) = \int \frac{\rmd^3k}{(2\pi)^3} \left| \tilde{W}(k \, r) \right|^2 P_{\rm L}(k),
 \label{eq:variance}
\eq
where $\tilde{W}$ is the Fourier transform of the real-space top-hat window function of radius $R$, i.e.,
\bq
 \tilde{W}(k \, R) = 3 \left[ \frac{\sin(k \, R)}{(k \, R)^3} - \frac{\cos(k \, R)}{(k \, R)^2} \right]
\eq
and $P_{\rm L}(k)$ is the linear matter power spectrum of the model.
We determine $P_{\rm L}(k)$ by using the Eisenstein-Hu transfer function~\cite{eisenstein:97a, eisenstein:97b} and the parametrized post-Friedmannian framework~\cite{hu:07b} for the designer $f(R)$ gravity model to determine the approximate growth needed to obtain the power spectrum (cf.~\cite{lombriser:10}).
The radius $r$ in Eq.~(\ref{eq:variance}) is defined by the cluster mass $M$ enclosed by it through
\bq
 r(M) = \left( \frac{3 M}{4\pi \, \brhom} \right)^{1/3},
 \label{eq:radiusmass}
\eq
which defines $\sigma_{\rm var}(M)$.

Given the virial overdensity $\Delta_{\rm vir}$ and a virial mass $\Mvir$, the virial radius $\rvir$ of the dark matter cluster is determined
by $\Mvir = 4\pi \, \brhom \, \Delta_{\rm vir} \rvir^3/3$.
We further assume a concentration $\cvir \equiv \rvir/\rs$  given by~\cite{bullock:99}
\bq
 \cvir(\Mvir,z=0) = 9 \left( \frac{M_*}{\Mvir} \right)^{0.13},
 \label{eq:concentrationscaling}
\eq
requiring $\sigma_{\rm var}(M_*)=\delta_c$.
This describes the characteristic scale $\rs$ as a function of the virial mass $\Mvir$.
The amplitude $\rhos$ of the NFW profile, Eq.~(\ref{eq:nfw}), is then defined by the integration to $\Mvir$.

The fits provided by this scaling relation are shown in Fig.~\ref{fig:coll}.
Note that the relation Eq.~(\ref{eq:concentrationscaling}) was obtained from fitting to halos of mass $(10^{11}-10^{14})~\Msunh$ produced in Newtonian CDM simulations~\cite{bullock:99} and the applicability to the halos in this study is therefore limited.
Nevertheless, this scaling relation is found to qualitatively reproduce the fitting parameters of the two extreme cases, i.e., $\Lambda$CDM and linearized $f(R)$ gravity.
The chameleon effect can be incorporated by interpolating the variance $\sigma_{\rm var}$ between its linearized $f(R)$ and $\Lambda$CDM value~\cite{li:11b}.
In~\cite{li:11b}, it was shown that with the help of such a transition function in $\sigma_{\rm var}$, one can describe the chameleon effect on the halo mass function and the nonlinear matter power spectrum.
Employing this interpolation in Eq.~(\ref{eq:concentrationscaling}), can change the value of $M_*$ in $f(R)$ gravity under the chameleon effect to its $\Lambda$CDM counterpart and hence, causes the concentration $\cvir$ to recover its $\Lambda$CDM limit for small values of $\absfR$.
This approach does, however, not include a dependency of the chameleon effect in $\cvir$ on the halo mass $\Mvir$, i.e., as long as utilizing the unaltered form of Eq.~(\ref{eq:concentrationscaling}).

Note that due to differences in the linear matter power spectrum between GR and the $f(R)$ model, the characteristic scale $\rs$ and therefore $\rhos$ for a given $\Mvir$ changes for $f(R)$ modifications with respect to the Newtonian results.
Hence, the relation $\rho(r)=\rho_{\rm GR}(r)$ (see~\textsection\ref{sec:selfsiminfallcollgas}) does not hold anymore.
Note, however, also that assuming modified forces in the gravitational collapse changes $\delta_{\rm c}$ and $\Delta_{\rm vir}$ and counteracts the changes in $\rhos$, $\rs$, and $\cvir$ (see~\cite{schmidt:08}) to some extent.
For the illustration in Fig.~\ref{fig:coll}, we assume standard forces in the derivation of $\delta_{\rm c}$ and $\Delta_{\rm vir}$, i.e., $F=0$.

Finally, in order to obtain the velocity dispersion at $\rs$, we revisit the Jeans equation, Eq.~(\ref{eq:jeans}), for a cluster produced with Newtonian gravity.
Assuming a NFW profile and the PPSD profile of Eq.~(\ref{eq:ppsd}), we obtain
\bq
 \sigma_{r, {\rm GR, s}}^2 = \kappa^2 \rho_{\rm GR, s} \rs^2 \frac{6 \ln 2-3}{25 - 24\beta_{\sigma, {\rm GR, s}}},
 \label{eq:sigmaspred}
\eq
where $\beta_{\sigma, {\rm GR, s}} \equiv \beta_{\sigma, {\rm GR}}(\rs)$.
We further assume that at $\rs$, the velocity anisotropy relation is correctly described by~\cite{hansen:04}
\bq
 \beta_{\sigma, {\rm GR, s}} \simeq \left. \frac{1}{40} \left( 17 - \frac{23}{6} \frac{\rmd \ln \rho}{\rmd \ln r} \right) \right|_{r = \rs} = \frac{37}{60},
 \label{eq:betaspred}
\eq
where $\rmd \ln \rho / \rmd \ln r = -2$ at $r=\rs$.
Eq.~(\ref{eq:betaspred}) was constructed as a fit to concordance model simulations in~\cite{hansen:04}.
The modified velocity dispersion is obtained from $\sigma_{r, {\rm s}}^2 = (1-F) \sigma^2_{r, {\rm GR, s}}$ with $F=-1/3$.
We assume $\beta_{\sigma, {\rm s}}=\beta_{\sigma, {\rm GR, s}}$ and
compare the predictions from Eqs.~(\ref{eq:sigmaspred}) and (\ref{eq:betaspred}) with the fits to the simulation output in Fig.~\ref{fig:coll}.
Note that $\sigmas^2 = \left( 3 - 2 \beta_{\sigma, {\rm s}} \right) \sigma_{r, {\rm s}}^2$.
This procedure yields a qualitatively good description of the velocity dispersions produced in the simulations.
Note that we do not assume a radial dependence of the force modification $F$ (cf. right panel of Fig.~\ref{fig:scalaron}) to determine the velocity dispersion in $f(R)$ gravity (see Fig.~\ref{fig:shape}) since the effect is blurred out over the different scales.
In Fig.~\ref{fig:coll}, we illustrate only the two extreme cases of the force modification, which correspond to the Newtonian and linearized $f(R)$ scenarios, respectively.
However, to estimate the correct halo-mass-dependent amplitude of the velocity dispersion, $\sigmas$, in the transition region to the chameleon regime, we can replace the constant force enhancement with a weighted average over the modification shown in the right panel of Fig.~\ref{fig:scalaron} (cf.~\cite{schmidt:10, clampitt:11}).

\vfill
\bibliographystyle{arxiv_physrev}
\bibliography{chameleonfR}

\end{document}